\documentclass[eqsecnum,preprint,flushrt]{aastex}

\usepackage{natbib}
\usepackage{graphicx}
\usepackage{gensymb}
\usepackage[small]{caption2}

\usepackage{amsmath, amsthm, amssymb }

\newcommand{\realL}{\mbox{$100$} Mm}
\newcommand{\realB}{\mbox{$10$} G}

\newcommand{\CSHalfLength}{\mbox{$100$} Mm}

\newcommand{\realTemp}{\mbox{$1$} MK}
\newcommand{\realnumdens}{\mbox{$10^{8}$} cm\textsuperscript{-3}}
\newcommand{\realbeta}{\mbox{$0.007$}}

\newcommand{\realReynolds}{\mbox{$2982$}}
\newcommand{\PrandtlSim}{\mbox{$0.01$}}

\begin{document}

\title{Density Enhancements and Voids following Patchy Reconnection}

\author{S. E. Guidoni, D. W. Longcope}

\affil{Department of Physics, Montana State University, Bozeman, MT 59717-3840, USA} 
\email{guidoni@physics.montana.edu}

\begin{abstract}

We show, through a simple patchy reconnection model, that retracting reconnected flux tubes may present elongated regions relatively devoid of plasma, as well as long lasting, dense central hot regions. Reconnection is assumed to happen in a small patch across a Syrovatski{\v i} (non-uniform) current sheet (CS) with skewed magnetic fields. The background magnetic pressure has its maximum at the center of the CS plane, and decreases toward the edges of the plane. The reconnection patch creates two V-shaped reconnected tubes that shorten as they retract in opposite directions, due to magnetic tension. One of them moves upward toward the top edge of the CS, and the other one moves downward toward the top of the underlying arcade. Rotational discontinuities (RDs) propagate along the legs of the tubes and generate parallel super-sonic flows that collide at the center of the tube. There, gas dynamics shocks that compress and heat the plasma are launched outwardly. The descending tube moves through the bottom part of the CS where it expands laterally in response to the background magnetic pressure. This effect may decrease plasma density by $30$ \%\ to $50$ \%\ of background levels. This tube will arrive at the top of the arcade that will slow it down to a stop. Here, the perpendicular dynamics is halted, but the parallel dynamics continues along its legs; the RDs are shut down, and the gas is rarified to even lower densities. The hot post-shock regions continue evolving, determining a long lasting hot region on top of the arcade. We provide an observational method based on total emission measure and mean temperature, that indicates where in the CS the tube has been reconnected.

\end{abstract}


%
\section{Introduction}

Reconnection, the mechanism where field lines are broken and connected to other ones with different topology, is present in many theories that undertake to explain the initiation and evolution of the biggest explosions in the solar system, flares or coronal mass ejections \citep{Lin_2000, Forbes_1996, Forbes_2006,Schrijver_2009}; see also \citep{Forbes_2000} for energy estimations in coronal mass ejections (CMEs)). Large amounts of stored magnetic energy can be released as thermal, kinetic, and radiative energy, as well as accelerated particles, when field lines change their topology \citep{Mandrini_2010}. Reconnection is also involved in the interaction of the solar wind with the Earth's magnetosphere, where substorms that may affect our space-weather, electronics and telecommunications, are originated. For a recent review on magnetic reconnection in space plasmas and laboratory, see \citet{Yamada_2010}. 

Direct observation of the actual reconnection process has been elusive. Its internal structure remains unknown, even for well-diagnosed laboratory experiments \citep{Katz_2010}. In order to explain the short times involved in the mentioned astrophysical explosions, fast reconnection \citep{Petschek_1964} ought to occur in a minuscule diffusion region on a CS where non-ideal field-line transport coefficients are locally enhanced. This  requires prohibitively high resolution measurements. Several fast reconnection models include localized magnetic resistivity enhancement on the CS \citep{Ugai_1977,Scholer_1987,Erkaev_2000,Ma_2001, Biskamp_2001}. 

All standard theoretical reconnection scenarios predict Alfv\'{e}nic outflows originated at the reconnection region. In steady state models, these outflows form coherent jets \citep{Parker_1957, Sweet_1958, Petschek_1964}, and in time-dependent models they are described as a localized retraction of field lines \citep{Semenov_1983, Biernat_1987,Heyn_1996,Nitta_2001}. 

Only in the last fifteen years or so, has technology started to close the gap between theory and observations. Nonetheless, only indirect measurements of solar corona reconnection have been possible. For example, the cusp-shaped, hot and dense flare loops \citep{Tsuneta_1992, Forbes_1996} consistent with the classical CSHKP model (\citet{Carmichael_1964, Sturrock_1968, Hirayama_1974, Kopp_1976}; see Figures 1 and 2 in \citet{Forbes_1996} for a modern version including thermal conduction) are considered observational evidence of reconnection. These arcades, slowly growing and whose footpoints separate as their height rises, are interpreted as closed magnetic loops formed after two-dimensional, steady state reconnection. They pile up on top of each other to form the arcades. The earlier loops cool down as new loops lay on top of them, and an apparent footpoint motion is observed. The loops are hot, as expected from the conversion of magnetic energy into thermal energy from reconnection. 

There are several other indirect observations of reconnection, like reconnection inflows \citep{Yokoyama_2001_I,Narukage_2006,Lin_2005} with speeds between a few km s\textsuperscript{-1} to approximately $106$ km s\textsuperscript{-1}, as well as Alfv\'{e}nic ($460-3500$ km s\textsuperscript{-1}) reconnection outflows \citep{Lin_2005,Wang_2007} and reconnection in/out pairs \citep{Sheeley_2007}. Loop shrinkage also seems to indicate reconnection. \citet{Forbes_1996}, using images of flare loops taken with SXT \citep{Tsuneta_1991} aboard Yohkoh \citep{Ogawara_1991}, showed reconnected cusped loops shrinking to form post-flare arcades. This process takes hours and is different from the fast passage ($1$-$2$ minutes) of field lines through the reconnection outflows (jets). Hard X-ray (HXR) sources near the apex of soft X-ray loops indicate some high-energy process, possibly electron acceleration in that region \citep{Masuda_1994,Krucker_2008,Tomczak_2009}. The acceleration may occur at the reconnection site or can be the result of the interaction of the downward reconnection outflow jet with the top of the SXR arcade, producing superhot plasmas \citep{Shibata_1995}. Plasmoid (magnetic island) formation \citep{Forbes_1983}, as a consequence of reconnection, has been widely observed (e.g. \citet{Shibata_1995,Nishizuka_2010}). For a comparative study of plasmoid in reconnecting CSs (solar and terrestrial contexts), see \citet{Lin_2008}. 

Even though most of the mentioned observations can be understood by standard, steady state and two-dimensional reconnection theories, there are newly observed phenomena that cannot be reconciled with these representations. Supra-arcade downflows (SADs) and supra-arcade downflowing loops (SADLs) seem to be two different observational signatures of retracting, isolated reconnected flux tubes with purely three-dimensional geometries \citep{Savage_2010}. 

\begin{figure}
   \centering
   \begin{center}$
      \begin{array}{cc}
         \resizebox{4.0in}{!}{\includegraphics*[74, 360][394, 735]{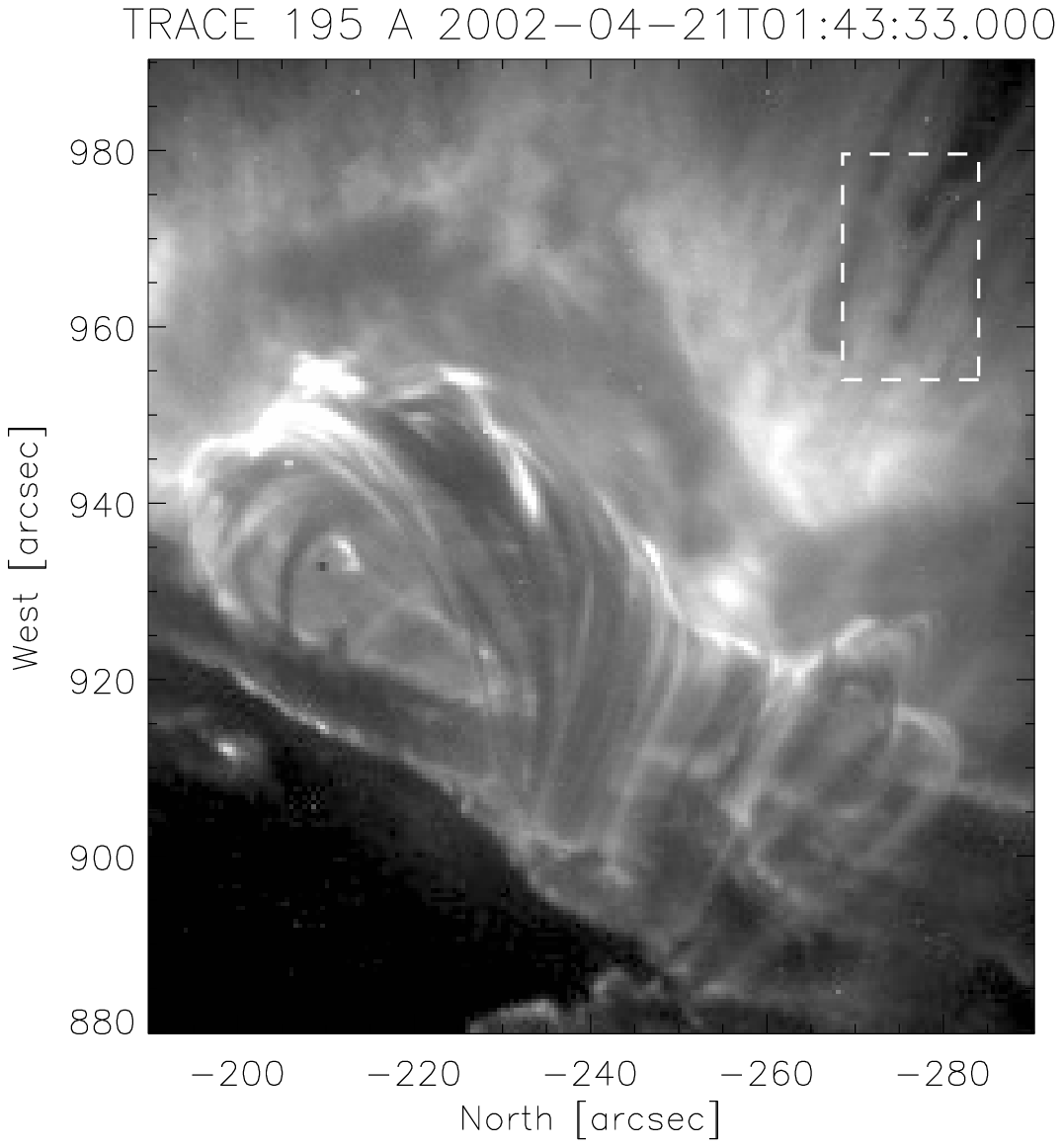}}
         \resizebox{!}{4.4in}{\includegraphics*[114, 350][370, 813]{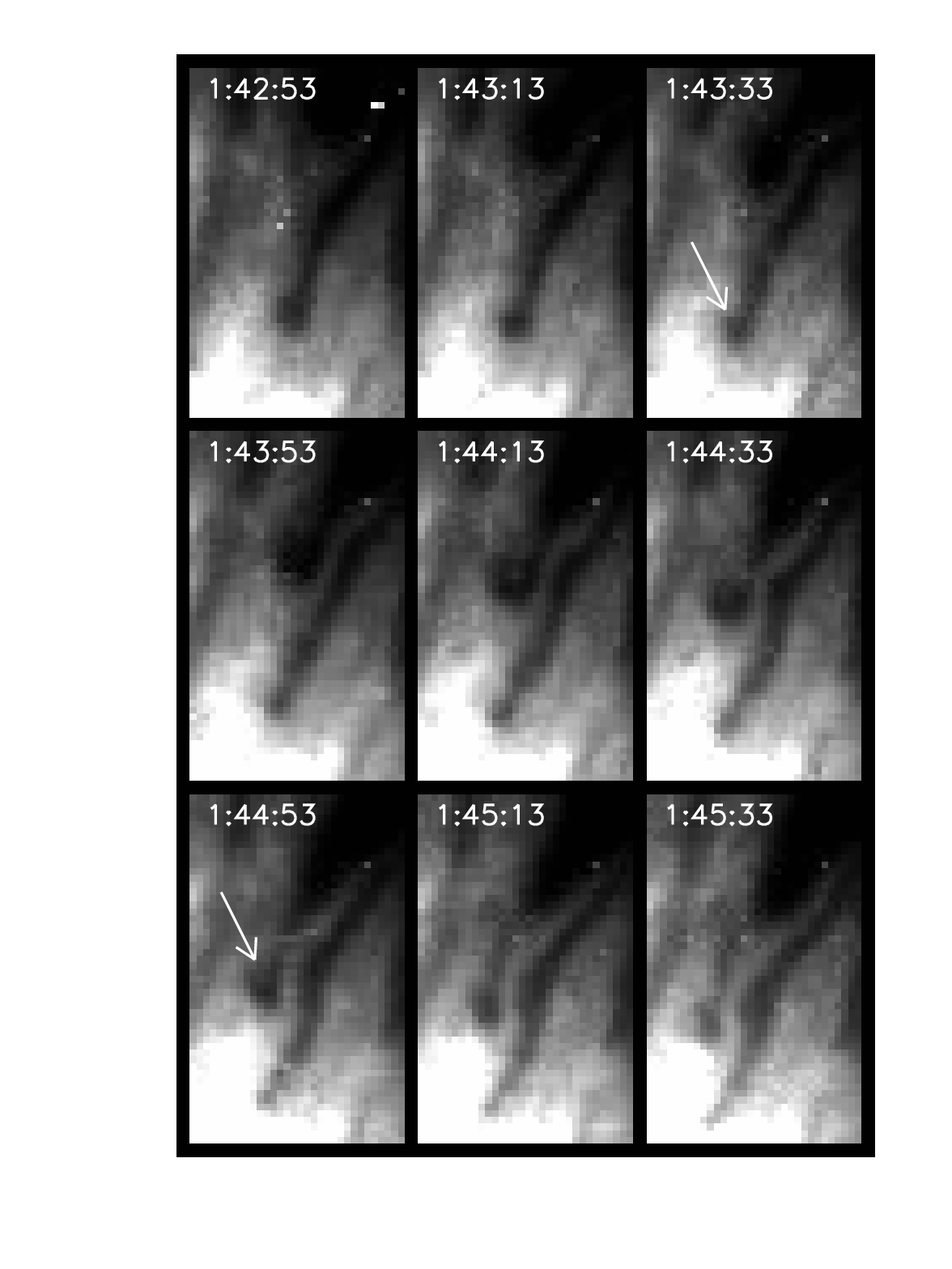}}
      \end{array}$
   \end{center}
   \caption { April 21, 2002 flare. Left panel: rotated TRACE 195 \AA\ image, taken at a time where dark voids can be seen descending toward the flare arcade. The dashed rectangle encloses a handful of them. Right panel: time sequence for the window delineated by the white dashed rectangle in the left panel. The white arrow in the 1:43:33 snapshot shows the position of one of the dark pockets. The other arrow points to a second localized void.}
   \label{fig:TRACE_SADs} 
\end{figure}

SADs were first reported more than one decade ago \citep{McKenzie_1999}, and have been continuously observed since then by several instruments \citep{McKenzie_2000, McKenzie_2001,Innes_2003, Innes_2003_II, Asai_2004, McKenzie_2009, Savage_2010}. They descend through the supra-arcade fan (see \citet{Svestka_1998} for a description of one of such fans) toward the underlying flare arcade and decelerate close to the arcade apex \citep{Sheeley_2002, Sheeley_2004}. They are spatially localized as dark pockets, followed by a dark wake or lane. Figure \ref{fig:TRACE_SADs} shows examples of SADs, as observed by TRACE \citep{Handy_1999,Schrijver_1999,Golub_1999} in 195 \AA\  passband. The image to the left shows an arcade for the April 21, 2002 X-class flare \citep{Wang_2002, Gallagher_2002, Innes_2003_II}, where two SADs are enclosed by a white dashed rectangle. Usually, arcades like this one can be described as two-dimensional, suggesting the presence of a wide spread CS above them. The images to the right are time sequences for the location delimited by the dashed rectangle. The white arrow on the top right snapshot indicates a dark pocket (SAD's head) that moves toward the arcade. This pocket is followed by a dark wake. The white arrow at time 1:44:33 indicates a second dark pocket (SAD) moving in the same direction. 

Downflows are observed as high as $40-60$ Mm above the top of soft X-ray arcades and interpreted as magnetic reconnection localized outflows \citep{McKenzie_2000}. Therefore, SADs appear to be the result of intermittent reconnection high in the corona. Three-dimensional and time-dependent models are needed to explain them. 

A three-dimensional model of patchy reconnection, accompanied by magnetohydrodynamic (MHD) simulations, has been presented by \citet{Linton_2006}. In that study, reconnection occurs across a one-dimensional, uniform CS with skewed magnetic fields. Resistivity is briefly enhanced by an unspecified mechanism within a sphere (reconnection patch). This creates a pair of bent, thin reconnected flux tubes that retract in opposite directions. RDs propagate along the legs of the tubes. Only the perpendicular dynamics of the tubes was analyzed by \citet{Linton_2006}, who also argued that the descending coronal voids are reconnected flux tubes descending from a flare site high in the corona. 

\citet{Longcope_2009} extended that study to include tube's parallel dynamics. The initial sharp angle at the reconnection site creates bends that move along the legs of the tubes as RDs. These bends rotate the magnetic field and generate parallel super-sonic inflows toward the center of the tube. There, the collision of these inflows launches outwardly two gas-dynamics shocks (GDSs) that heat and compress the plasma. The GDSs are disconnected from the resistive region, and are features of the ideal relaxation following reconnection. The rate of reconnection is not relevant in this case since the diffusion region is short-lived and does not participate in the post-reconnection evolution. 

In a recent paper \citep{Guidoni_2010}, we presented generalized thin flux tube (TFT) equations for the dynamics of reconnected flux tubes. These include a pressure gradient in the parallel direction, as well as temperature-dependent, anisotropic viscosity and thermal conductivity in non-ideal terms. In that paper, the evolution of reconnected tubes across a uniform CS and skewed magnetic fields is analyzed in detail through simulations and modeling. The inner structure of the GDSs consists of a thermal front where the temperature increases and most of the heating occurs, followed by an isothermal sub-shock where plasma is compressed. The thickness of the shocks, comparable to the entire length of the tubes, is calculated theoretically.

Patchy reconnection seems to agree with several observations of reconnection. For example, \citet{Longcope_2005}, through EUV observations and modeling, analyzed an emerging active region in the vicinity of an existing one. They studied non-flaring reconnection, along a separator overlying the volume between them. They characterized $43$ loops observed by TRACE that result from reconnection between the active regions, and assumed a correspondence between them and model flux tubes. The average diameter of the loops is $3.7$ Mm with very small variation, suggesting a sporadic or patchy reconnection. The estimated flux of each one of them is $~4 \times 10^{18}$ Mx for an assumed magnetic field of $37$ G. Figure 19 of the paper shows an illustration of the reconnection model applied to the active regions where the three-dimensional CS is shown, as well as a pair of reconnected flux tubes created by patches of reconnection.  The time history of the 171 \AA\ loops with very similar diameters suggests a highly intermittent reconnection process. 

What triggers reconnection is still an open question, but several promising mechanisms have been proposed. For example, the MHD CS tearing mode instability \citep{Furth_1963,Sturrock_1966,Forbes_1983,Priest_1985, Biskamp_1986} may initiate bursty reconnection. \citet{Shimizu_2009} showed numerically that a two-dimensional fast magnetic reconnection in a simple one-dimensional CS can be unstable for three-dimensional resistive perturbations. Therefore, it is possible to randomly eject three-dimensional magnetic loops along a CS atop a post-flare arcade, resembling SADs. Catastrophe models explain how magnetic configurations are stable for a long time and suddenly release their energy \citep{Cassak_2005}. The Hall reconnection model, extensively studied in the Geospace Environmental Modeling(GEM) magnetic reconnection challenge (\citet{Birn_2001}, and companion papers), yields energy release rates consistent with the ones observed in the solar corona \citep{Drake_2006_I}.

Even though three-dimensional outflows can be generated from purely two-dimensional configurations, the assumption of perfectly anti-parallel pre-reconnection magnetic fields seems too restrictive for the complex solar corona, where field lines are likely to form finite angles across CSs.  Furthermore, the addition of a guide field is directly related to the temporal evolution of reconnection. \citet{Drake_2006_I} shows that ambient guide magnetic field controls whether magnetic reconnection, once it is established, remains steady (anti-parallel reconnection) or becomes bursty (reconnection with a guide field). Through particle simulations in a system consisting on two Harris CSs, they show that magnetic reconnection in a large-scale system with a guide field will break up into many islands, as a result of tearing instability.  Guide fields with a shear angle of $127$\degree\ are sufficient to produce secondary islands that grow to finite amplitude, producing well developed flux tubes. 

When the standard reconnection Petschek model is extended to include skewed fields, the original V-shaped ``switch-off'' shocks are replaced by RDs (where the field direction is changed and plasma is accelerated) and slow-shocks (SSs; where kinetic and magnetic energy are partially converted to heat.) \citep{Petschek_1967,Soward_1982,Skender_2003}; see Figure 1 in \citet{Longcope_2010_I} for a single flux tube as it passes through these shocks). Therefore, the energy conversion occurs in a two-step process. For the reconnected tubes by \citet{Longcope_2009} and \citet{Guidoni_2010}, the energy conversion also has similar behavior: first, magnetic energy is converted to kinetic energy at the RDs, and later this kinetic energy is partially converted to thermal energy at the GDSs. GDSs are parallel shocks, closely related to SSs. 

Besides the assumption of being retracting reconnected flux tubes, there are several unanswered questions about SADs. To the best of our knowledge, there has not been a satisfactory explanation of why they seem to be hot and relatively devoid of plasma. \citet{McKenzie_1999} reported temperature ratios between the voids and the surrounding rays of the fan of $7.9$ to $9.1$. The wakes seem to have a slightly lower ratio: $8.6$ to $9.1$. The density ratios between the voids and the rays ranged between $1.3/2.3$ to $1.0/4.0$, depending on the assumptions about the line of sight depth. The darkness of the voids in X-ray and extreme-ultraviolet (EUV) images and spectra, together with the lack of absorption signatures in EUV \citet{Innes_2003} are best explained as pockets of very low plasma density. \citet{Innes_2003} ruled out the possibility that the dark tracks are caused by coronal rain (cold plasma falling gravitationally). 

In the present paper, through a simple but realistic fluid model of patchy reconnection, we show that plasma depletion naturally occurs in flux tubes that are reconnected across Syrovatski{\v i}-type CSs \citep{Syrovatskii_1971}. In this kind of CSs, the background magnetic pressure has its maximum at the center of the CS plane, and decreases toward the edges of the plane. While the tube descends toward the top of the arcade, plasma density is decreased by tube's lateral expansions in response to the external background pressure. After the tubes lay on top of the arcade, rarefaction waves continue decreasing plasma density. 

Syrovatski{\v i}-type CSs may be ubiquitous in the solar corona. \citet{Edmondson_2010} show that even a simple multipole topology would form very stable CSs bounded by two Y-type nulls, where an X-type null deforms under stresses due to motions in the photosphere. They did not include a guide field, but suggested that if added, the expected three-dimensional islands formed after reconnection may resemble longer flux-tube-like magnetic islands. 

There is another piece of the puzzle that has not been completely understood: SADs' reported speeds (30 -  500 km s\textsuperscript{-1}) are smaller than the assumed Alfv\'{e}n speed in the corona \citep{McKenzie_1999,McKenzie_2000, Asai_2004, McKenzie_2009, Savage_2010}, and this contradicts any standard reconnection theory. A possible explanation for reconnected tube's deceleration is given by \citet{Linton_2006}. They showed that a drag effect from the plasma around the tube can reduce its speed considerably. The external fluid deforms and accelerates as the tube moves through it; this requires energy and therefore slows down the tube. The ratio between the speed observed in their simulations and the perpendicular Alfv\'{e}n speed (expected speed from the TFT model) increases with the reconnection angle and ranges approximately between $0.1$ and $0.5$ for the low plasma-$\beta$ (ratio between thermal pressure and magnetic pressure) case. This ``added mass'' effect accounts for most of the flux tube drag at high plasma-$\beta$, but is not enough, however, to explain the speed reduction at low plasma-$\beta$ (the expected speeds taking into account added mass effects are double the one observed in the simulations). They conclude that there is an additional source of drag in this case, but they do not speculate about its origin. 

We do not know the correct answer to the disparity between observations and models regarding the speed of localized reconnection outflows, but most likely the slowing down is due to the interaction between tubes and their surroundings. In our present work, for simplicity, this interaction is not included; the reconnected tubes are assumed to be completely isolated from their background. Therefore, we restrict ourselves to standard reconnection scenarios where the outflows are Alfv\'{e}nic. 

The layout of this paper is as following. The first section describes the background configuration where reconnection is assumed to happen. Three different locations of reconnection across the CS are described. The next section presents the TFT equations that govern the evolution of the reconnected flux tubes, as well as analytical equilibrium solutions to the perpendicular part of the equations. Section \ref{sec:simul} describes simulations of the evolution of reconnected tubes for each reconnection location. The parallel dynamics of the tube is described in detail before the tube arrives at the top of the underlying arcade. In Section \ref{sec:Top_arcade}, the arrival at the arcade apex is simulated by an overdamped spring force exerted by the compressed arcade that slows down and stops the tube. After this arrival, the parallel dynamics changes and strong rarefactions develop. In Section \ref{sec:rarefaction}, the temporal evolution of the tube's total emission measure and mean temperature are shown, and a method is described to determine where in the CS a tube was reconnected. The last section presents a discussion of our results and their possible observational consequences.

%
%
%
\section{Flare Current Sheet in the Solar Corona}
  \label{sec:Green_Syrovatskii}

For the location of the reconnection episode in the solar corona, we assume a realistic magnetic field configuration like the one shown in Figure \ref{fig:flare_CS}. This initial equilibrium background presents a CS in the $x-y$ plane, located high in the corona; and skewed magnetic fields on each of its sides. The mathematical description of this configuration is usually called Green-Syrovatski{\v i} \citep{Green_1965,Syrovatskii_1971} CS, or ``double y-type'' due to the shape of its separatrices when projecting field lines in the $y-z$ plane (see panel (b) of Figure \ref{fig:flare_CS} in the present paper, and Figure 3.b in Syrovatski{\v i}'s paper). We included a constant guide field in the ignorable $x$ direction to provide an arbitrary angle between the pre-reconnection field lines. We also assume uniform initial values of density $\rho_{e}$, and pressure $P_{e}$, on each side of the CS. The CS extends infinitely in the $x$-direction, but is finite in the $y$-direction. The bottom edge of the CS marks the top of the flare arcade (shown in Figure \ref{fig:flare_CS} as arc-shaped loops) with footpoints in the solar surface (bottom panel of the box). We will denote the half length of the CS in the $y$-direction as $L_{e}$. 

The entire region is assumed to be free of electrical resistivity, with exception of a short-lived small patch, somewhere in the CS where reconnection occurs (small sphere, in Figure \ref{fig:flare_CS}). The horizontal and vertical position of the reconnection site will be denoted by $x_{R}$ and $y_{R}$, respectively. Any horizontal position of the reconnection site is equivalent due to the initial 2 1/2 dimensional symmetry of the magnetic configuration described above. For simplicity, we will assume it to be at the origin of the horizontal axis, then $x_{R} = 0$. 

We will distinguish three reconnection cases, based on their location on the CS. If reconnection happens in the top half of the CS ($0 < y_{R} < L_{e}$), it will be called TOP reconnection, if it happens in the middle of the CS ($y_{R}=0$), it will be denoted CENTER reconnection, and finally if reconnection occurs somewhere in the bottom half of the CS, it will be referred as BOTTOM reconnection. Figure \ref{fig:CS_lines} represents the region of the CS enclosed by a dashed rectangle in Figure \ref{fig:flare_CS} and shows examples of the three different reconnection positions. For the sake of concreteness, we have assumed that the half length of the CS is $L_{e} = $ \realL\ , a typical length in the solar corona.  

The magnetic field near the CS can be described as the limit of $z \rightarrow 0$ of the skewed Green-Syrovatski{\v i} magnetic field, as follows 
%
\begin{eqnarray}
   \label{eqn:B_syrovatskii}
     \mathbf{B_{e}} &=& \widehat{\mathbf{x}} B_{ex} \pm \widehat{\mathbf{y}} B_{ey} \sqrt{1-\left(\frac{y}{L_{e}}\right)^{2}}  \nonumber \\ 
       &=& \widehat{\mathbf{x}} B_{ex}  \pm \widehat{\mathbf{y}} B_{ey} \sqrt{1-y^{\prime 2}}  ,\hbox{    for } z \gtrless 0.
\end{eqnarray}
%
Here, $B_{ex}$ and $B_{ey}$ are constants. $B_{ex}$ is the guide field, and $B_{ey}$ is the $y$-component of the usual Green-Syrovatski{\v i} field. The positive (negative) sign corresponds to the front (back) side of the CS. For simplicity, we defined unitless prime-variables, $y^{\prime} = y/L_{e}$ and $x^{\prime} = x/L_{e}$.  The lines $y^{\prime} = \pm 1$ represent the edges of the CS. The magnitude of the external field is maximum at the center of the CS and decreases toward the edges, as so does the corresponding local Alfv\'{e}n speed.

Only field lines in the vicinity of the plane of the CS that intersect the resistive patch will be reconnected, forming two opposite thin flux tubes. In Figure \ref{fig:CS_lines}, spherical resistive patches are represented by their circular projections in the plane of the CS. Reconnected flux tubes have a finite thickness, as also does the CS. The tube's cross sectional area at the reconnection region, $A$, is determined by the radius of the reconnection patch. This area and the magnitude of the magnetic field at the reconnection site determine the flux of the tube, $\phi = B A$. 

The actual size of the reconnection patches in the solar corona has not been determined, but if SADs represent the cross sectional area of reconnected flux tubes, this value can be estimated to be only a few megameters in radius, as shown in Figure \ref{fig:TRACE_SADs}. \citet{McKenzie_2009} reported dark voids sizes of approximately $10^{7}$  km\textsuperscript{2}, with significant range for variation.  The areas of the dark pockets can only give an estimation because they may change as tubes move, due to pressure balance with the background. Typical flare coronal loops extend for a hundred megameters, therefore it is reasonable to assume that the reconnected tubes are thin compared with their length.

Field lines in front of the CS that intersect a given reconnection region will connect at this site with field lines on the other side of the CS. Figure \ref{fig:CS_lines} shows a sample of three pre-reconnection field lines on each side of the CS that intersect a TOP reconnection site located at $y_{R}^{\prime}=3/4$. We will assume that the pre-reconnection flux tubes are parameterized by one central field line (their axes, thicker lines in the graph), that crosses the center of the reconnection region. In this figure, only pre-reconnection representative field lines are drawn for the CENTER case. 

The half angle $\zeta_{R}$ between the pre-reconnection representative field lines at a given reconnection location is a parameter in our model, and can be calculated as
%
\begin{eqnarray}
   \label{eqn:B_rec_angle}
     \tan{\left(\zeta_{R}\right)} = \frac{B_{ey}}{B_{ex}} \sqrt{1-y_{R}^{\prime 2}} = \tan{\left(\zeta_{0}\right)} \sqrt{1-y_{R}^{\prime 2}},
\end{eqnarray}
%
where $\zeta_{0}$ represents the half angle that field lines make at the center of the CS. Both angles are shown in Figure \ref{fig:CS_lines}. 

Any pre-reconnection field line near the CS can be described as 
%
\begin{eqnarray}
   \label{eqn:initial_tube}
      y^{\prime}(x^{\prime}) &=& \pm \sin \left[ \tan{\left(\zeta_{0}\right)} x^{\prime} + c \right], \hbox{  } z \gtrless 0,
\end{eqnarray}
%
with $c$ being a constant that can be determined by any point that intersects the curve. For field lines that cross the reconnection point $(0,y_{R}^{\prime})$, $c = \pm \arcsin(y_{R}^{\prime})$, where the top sign corresponds to field lines in front of the CS ($z > 0$, thick solid lines in Figure \ref{fig:CS_lines}), and the bottom sign corresponds to field lines at the back of the CS ($z < 0$, thick dotted lines in Figure \ref{fig:CS_lines}). 

The $x$-position at which a tube (its representative field line) is the closest to an edge of the CS can be found replacing $y^{\prime} = \pm 1$ in Equation \eqref{eqn:initial_tube} (the positive sign corresponds to the top edge of the CS, and the negative sign corresponds to bottom edge of the CS), to obtain
%
\begin{eqnarray}
   \label{eqn:end_points}
      x_{top}^{\prime} &=& \pm \left[\frac{\frac{\pi}{2} - \arcsin \left({y_{R}^{\prime}} \right)}{\tan{\left(\zeta_{0}\right)}} \right] , \hbox{  } z \gtrless 0, \nonumber \\
      x_{bottom}^{\prime} &=& \mp \left[\frac{\frac{\pi}{2} + \arcsin \left({y_{R}^{\prime}} \right)}{\tan{\left(\zeta_{0}\right)}} \right] , \hbox{  } z \gtrless 0.
\end{eqnarray}
%
At these locations, the tubes are tangent to the edge of the CS since the only remaining component of the magnetic field is the guide field (Equation \eqref{eqn:B_syrovatskii}). The tubes have legs that continue downward to the solar surface and these sections of the tubes do not generally lay in the same plane as the CS (see Figure \ref{fig:flare_CS}). 

After reconnection, the reconnected flux tubes are symmetric in the $x$-direction, but the initial $2$ $1/2$-d symmetry of the system is lost. The localization of the reconnection episode makes the problem three-dimensional. Fortunately, the reconnected tubes are mostly two-dimensional as they lay in the plane of the CS, and their retraction is along this plane (magnetic tension force is parallel to the plane of the CS). In Figure \ref{fig:CS_lines}, representative lines for reconnected flux tubes in a BOTTOM reconnection site are shown. The dashed line represents the initial configuration for a reconnected flux tube whose left side was part of a tube on the back side of the CS and whose right side was part of a tube in the front side of the CS. This new tube is sharply bent near the reconnection site and will move upward as a transient feature that slides along the CS, between the flux layers. We call these tubes UPWARD moving tubes. The mixed dotted and dashed line in the figure is the initial configuration of the DOWNWARD moving reconnected tube that will retract in the negative $y$-direction. The directions of motion for each tube are indicated in the figure by arrows pointing outward from the BOTTOM reconnection site.  The UPWARD and DOWNWARD moving tube definitions are valid for all reconnection cases. Any reconnected UPWARD moving tube is a mirror image of a DOWNWARD moving tube if both tubes have the same absolute value $y_{R}^{\prime}$ reconnection location, but with opposite sign. They will move in opposite directions, but their dynamics and shapes are equal. Therefore, in this paper we will only focus on DOWNWARD moving tubes. 

The shape of a reconnected tube changes with time mainly as result of magnetic tension. This motion can be characterized by its representative field line arc-length parametrization, $\mathbf{R}(l,t)$. At each point of the tube, the local tangent unit vector and curvature vector can be calculated from the parametrization as $\widehat{\mathbf{b}} = \partial \mathbf{R} /\partial l$ and $\mathbf{k}= \partial \widehat{\mathbf{b}} /\partial l$, respectively. 
Figure \ref{fig:CS_lines} shows an example of the parametrization vector for an initial UPWARD moving tube, at a generic arc-length $l$. Also, at this point of the curve, $\widehat{\mathbf{b}}$ and $\mathbf{k}$ are shown. Pieces of the tube like this one, that are far enough from the reconnection site, are initially in equilibrium since they were part of the original background configuration. There, the force due to the curvature of the tube is balanced by the force due to the external gradient of the magnetic field. 

%
%
%
\section{Low Plasma-$\beta$ Thin Flux Tube Equations}
  \label{sec:TFT_Syr}

Reconnection is assumed to happen in a small and short-lived region, therefore the reconnected tubes are thin and their evolution can be described by the TFT equations presented by \citet{Guidoni_2010}. These equations assume total pressure balance between the tubes and their surrounding plasma, and the tubes are assumed to be isolated from the background (no viscous momentum and heat exchange). The plasma in these tubes satisfies the frozen-in field condition since the resistivity that may have originated the reconnection episode was only non-zero at the reconnection region. 

In the solar corona, the plasma is strongly magnetized, ergo its plasma-$\beta$, is small. For this reason, thermal pressure can be neglected with respect to magnetic pressure. This approximation and the assumption of total pressure balance constrain the magnitude of the magnetic field inside the tube; at all times, it is equal to the magnitude of the external magnetic field $B_{e}$ \citep{Guidoni_2010}, and can be calculated from Equation \ref{eqn:B_syrovatskii}. 

The cross sectional area of each element of the tube changes as the element moves between the flux layers of the CS, satisfying $A = \frac{\phi}{B_{e}}$. The density of each tube element of mass $\delta m$ varies according to $\rho = \delta m B_{e} / \phi \delta l$, where $\delta l$ is the length of the element. 

The magnetic field direction is completely described by the positions of the tube elements since the frozen-in condition guarantees that fluid particles and field lines move together. These positions can be obtained integrating the velocity satisfying the low plasma-$\beta$ TFT momentum equation \citep{Guidoni_2010}
%
\begin{eqnarray}
   \label{eqn:TFT_mom} 
      \rho \frac{D\mathbf{v}}{Dt} & = &  -\widehat{\mathbf{b}} \frac{\partial P}{\partial l}-\nabla_{\perp}
      \left(\frac{B_{e}^{2}}{8 \pi}\right ) + \mathbf{k}\left(\frac{B_{e}^{2}}{4 \pi}\right) +  B_{e} \frac{\partial}{\partial l} \left[ \frac{\widehat{\mathbf{b}}\eta}{B_{e}}\left(\widehat{\mathbf{b}}\cdot \frac{\partial \mathbf{v}}{\partial l}\right)\right]. 
\end{eqnarray}
%
%
\noindent Here, $\rho$, $\mathbf{v}$, $P$, and $\eta$ are the mass density, velocity, plasma pressure, and viscosity of the fluid inside the tube, respectively. $D/Dt$ is the advective derivative in the tube element's reference frame and $\nabla_{\perp}$ represents the gradient in the direction perpendicular to the flux tube. 

The inviscid part of the above equation has a parallel-to-the-magnetic-field term with a derivative of the thermal pressure that cannot be neglected with respect to any magnetic pressure term since magnetic forces act only in the perpendicular direction (as shown in the second and third term of the right-hand side). The viscosity term has components in both directions but is usually small since viscous Reynolds numbers in the corona are very large. The exception to this rule are shocks, where velocity gradients become significant. For strongly magnetized plasmas, the viscosity depends on temperature as $\eta = \eta_{c}T^{\frac{5}{2}}$ with $\eta_{c}$ being a constant. 

The contribution of thermal conductivity to the change in entropy of the plasma is approximately two orders of magnitude larger than the viscosity. The change in entropy of the plasma (assumed to be an ideal gas) of any given tube element can be expressed as $\Delta s = \ln{P_{c}/P_{e}}$, where $P_{c}$ is related to the tube element's pressure and density as 
%
\begin{eqnarray}
    \label{eqn:TFT_Pc}
      P(t) & = & P_{c}(t) \left(\frac{\rho(t)}{\rho_{e}} \right)^{\gamma}. 
\end{eqnarray}
Here, $\gamma$ is the adiabatic gas constant.
%
%

The TFT entropy equation can be transformed in an equation for $P_{c}$, as follows \citep{Guidoni_2010}
%
%
\begin{eqnarray}
   \label{eqn:TFT_Pc_time}
      \frac{D P_{c}}{Dt} & = & (\gamma -1)  \left( \frac{\rho_{e}}{\rho } \right)^{\gamma} 
      \left[ \eta \left(\widehat{\mathbf{b}} \cdot \frac{\partial \mathbf{v}}{\partial l} 
      \right)^{2} + B_{e} \frac{\partial}{\partial l} \left( \frac{\kappa}{B_{e}}   \frac{\partial}{\partial l}  (k_{B} T) 
      \right) \right].
\end{eqnarray}
%
The temperature dependence of the thermal conductivity is the same as for the viscosity, $\kappa = \kappa_{c}T^{\frac{5}{2}}$, with $\kappa_{c}$ being a constant. The ideal gas constitutive relation is $P = \rho k_{B} T/ \overline{m} = 2 n k_{B} T$, with $k_{B}$ being the Boltzmann constant, $\overline{m}$ the average particle mass, and $n$ the electron number density.

The TFT mass conservation has the following expression \citep{Guidoni_2010} 
%
%
\begin{eqnarray}
   \label{eqn:TFT_mass} 
      \frac{D}{Dt}\left(\frac{B_{e}}{\rho}\right) & = & \frac{B_{e}}{\rho} \widehat{\mathbf{b}} \cdot \frac{\partial \mathbf{v}}{\partial l}.
\end{eqnarray}
%

One of the goals of this paper is to explain how reconnected flux tubes may become regions of density depletion, like in the dark voids mentioned in the Introduction. To that end, the above equation can be re-arranged to show that the density of each tube element may change due to three different effects, 
%
%
\begin{eqnarray}
   \label{eqn:density change} 
      \frac{D}{Dt}\ln \left(\rho \right) & = & -\frac{D}{Dt}\ln \left( A \right) - \frac{\partial v_{\parallel}}{\partial l} + \mathbf{v}_{\perp} \cdot \mathbf{k},
\end{eqnarray}
%
where $v_{\parallel}$ is the parallel component of the velocity and $ \mathbf{v}_{\perp} $ is the perpendicular one. 

The first term in the right-hand side corresponds to a change in area of the tube as the tube element moves to regions of different confining magnetic field magnitude. In addition, if neighboring elements are slowing down ahead of the direction of motion, a pile-up occurs and the density of the element increases (its length gets reduced, and its area is fixed by the external field). The opposite effect occurs if the neighboring elements are slowing down upstream. The second term in the right-hand side reflects this effect, which is similar to what happens in traffic jams. The third term in the right-hand side is related to the expansion or contraction of a curved tube in its perpendicular direction. One example of this effect is a circular tube with fixed area expanding or contracting radially. In this case, the curvature vector always points toward the center of the tube, and the perpendicular velocity is parallel (contraction) or antiparallel (expansion) to this vector. Each tube element expands or contracts resulting in a change in density. We will show the relevance of some of these terms in section \ref{sec:parall_dyn}.


\subsection{Joined Equilibrium Solution}
  \label{sec:JES}

A simpler magnetic background configuration than the one assumed in the present paper was studied by \citet{Guidoni_2010}. There, the background field is skewed and uniform, and the shape of the reconnected thin flux tubes, as they evolve, can be described by a joined equilibrium solution (JES). We will call this uniform background model \textit{Guidoni10}. In this case, the reconnected tubes, at any instant, are composed by three straight segments joined at two corners (the bends) that move at the Alfv\'{e}n speed along the initial tube. There, the magnetic field is rotated without changing its magnitude (rotational discontinuity). Each straight segment of the tube is an equilibrium solution of the inviscid perpendicular part of the momentum TFT equation, but the joined solution is not in equilibrium due to the sharp angle at the bends. The two lateral segments are at rest and the central segment moves at the projection of the Alfv\'{e}n speed in the y-direction. Since the magnetic field magnitude and the initial density are uniform, the Alfv\'{e}n speed is the same everywhere. 

The actual parallel dynamics in Guidoni10 does not satisfy an equilibrium solution (constant pressure and temperature) since strong GDSs develop inside the tubes. Nevertheless, outside the shocks, the density and temperature are approximately constant. To completely satisfy an equilibrium solution of the general TFT equations, the change in velocity should satisfy $\widehat{\mathbf{b}} \cdot \frac{\partial \mathbf{v}}{\partial l}  =  0$; a condition that is not satisfied at the shocks. 

For our current non-uniform magnetic configuration (Equation \eqref{eqn:B_syrovatskii}), it is also possible to find an analytical solution for the inviscid perpendicular part of the momentum TFT Equation \eqref{eqn:TFT_mom}, where each mass element is in equilibrium ($ \frac{D }{Dt} = 0 $). In this case, the analytical solution differs quantitatively with the actual tube dynamics, as we will see in the next section, but the general aspects of the solution are maintained.  

It is straightforward to see that a tube satisfying the following equation constitutes such an equilibrium solution
%
%
\begin{eqnarray}
   \label{eqn:Equil_B} 
      \mathbf{k} & = & \frac{\partial^{2} \mathbf{R}} {\partial l^{2}} = \frac{1}{2} \nabla_{\perp} \ln{\left( B_{e}^{2} \right)} . 
\end{eqnarray}
%
A possible general equilibrium solution of the TFT equations is given by a tube having uniform density and temperature that satisfies the above equation, as well as $\widehat{\mathbf{b}} \cdot \frac{\partial \mathbf{v}}{\partial l}  =  0$. In this section, we will only focus on the shape of the tube given by
\eqref{eqn:Equil_B}, as the parallel dynamics is not expected to be in equilibrium. 

The reconnected tubes lay in the plane of the CS, then the derivative with respect of the arc-length of the tube can be expressed as $\partial/ \partial l = (1 / \sqrt{1+ \dot x^{\prime 2}}) \partial/ \partial y^{\prime}$, where $\dot x^{\prime} = \partial x^{\prime}  / \partial y^{\prime}$. With this expression, after some algebra, Equation \eqref{eqn:Equil_B} becomes 
%
\begin{eqnarray}
   \label{eqn:Equil_der} 
      \frac{\partial}{\partial l} \ln{B_{e}^2} & = & \frac{- 2 \ddot x^{\prime}}{\dot x^{\prime} (1 + \dot x^{\prime 2})} = \frac{\partial}{\partial y^{\prime}} \ln{\left(\frac{1 + \dot x^{\prime 2}}{\dot x^{\prime 2}}\right)}.
\end{eqnarray}
%
The initial background configuration satisfies the general MHD equation's equilibrium. Its field lines \eqref{eqn:initial_tube}, also satisfy the above equation. Therefore, pre-reconnection tubes are in equilibrium with respect to the general TFT equations. Immediately after reconnection, every tube element is also in equilibrium, except the central element that is sharply bent at the reconnection site.

The reconnected flux tubes are symmetric in the x direction, and smooth at their center. A solution to Equation \ref{eqn:Equil_der} that satisfies these conditions can be expressed as
%
\begin{eqnarray}
   \label{eqn:Equil_sol} 
      y^{\prime}(x^{\prime}) & = & y_{c}^{\prime} \cos{\left[\frac{\tan{\left(\zeta_{0}\right)} x^{\prime}}  {  \sqrt{   1 + \left(1 - y_{c}^{\prime 2}\right)\tan^{2}{\left (\zeta_{0} \right)} } } \right]}.
\end{eqnarray}
%
This solution is a one-parameter family of curves whose parameter is the $y$-position of the center of the tube, $y^{\prime}_{c}$.

It is possible to construct a JES where the shape of the reconnected tubes is composed of three different equilibrium regions: two unperturbed equilibrium side sections given by Equation \eqref{eqn:initial_tube}, and a central curve described by the above equation. The left panel of Figure \ref{fig:CENTER_JES_sim} shows such solutions for a CENTER reconnection case with half reconnection angle $\zeta_{R}=\zeta_{0}=45$\degree, and DOWNWARD moving tube. Each dotted line corresponds to a different parameter (position of the center of the tube, represented in the figure by triangles). The unperturbed side sections are part of the initial reconnected tube (solid lines in the figure). For example, a complete JES for the parameter $y^{\prime}_{c}$ labeled ``C'' in the figure, corresponds to two side sections of the initial tube (sections ``AB'' and ``DE''), plus a central symmetric curve (``BCD''). For a given center position, the intersection between the dotted curve and the solid line (the bends, shown as circles in the figure) can be obtained setting Equations \eqref{eqn:Equil_sol} and \eqref{eqn:initial_tube} equal to each other. 

The whole set of these JESs describes a reconnected tube that descends from the reconnection region. At each time, the tube is shorter than the original tube. Even though each section of the joined solution is an equilibrium solution, the joined solution is not. The bends are clearly out of equilibrium, therefore a complete curve like ``ABCDE'' does not satisfy equilibrium Equation \eqref{eqn:Equil_der}. Left panels of Figures \ref{fig:BOTTOM_JES_sim} and \ref{fig:TOP_JES_sim} show similar joined equilibrium solutions for BOTTOM and TOP reconnection cases, respectively. 

For comparison purposes, all cases analyzed in the present section have a half reconnection angle $\zeta_{R}=45$\degree\ (note that this is different for the CENTER reconnection case shown in Figure \ref{fig:CS_lines} that has $\zeta_{R} \simeq 57$\degree; this choice visually simplified the graph making pre-reconnection field lines parallel to each other on each side of the CS, for all the reconnection cases). From now on, CENTER reconnection case implies $\zeta_{R}=45$\degree. 

In the CENTER and BOTTOM reconnection cases, the concavity for the JESs is up for all the central regions. On the other hand, for the TOP reconnection case, the concavity of the central regions changes from down to up when the center position changes from positive to negative (at $y^{\prime}_{c} = 0$, the equilibrium central region corresponds to a straight horizontal line). 

Equilibrium solutions \eqref{eqn:Equil_sol} are valid as long as the position of the center of the tube is above the edge of the CS ($y^{\prime}_{c} > -1$, or $y_{c} > $ \CSHalfLength). This corresponds to the arrival of the center of the tube to the top of the arcade. Since our analysis was restricted to field lines near the plane of the CS, equilibrium solutions beyond this point have no real significance.

%
%
%
\section{Simulations}
  \label{sec:simul}

The low plasma-$\beta$ TFT Equations \eqref{eqn:TFT_mom}, \eqref{eqn:TFT_Pc_time}, and \eqref{eqn:TFT_mass} are non-linear, hence a time-dependent analytical solution of these equations is hardly a possibility. The JES, although far from being a general solution, presents the correct general aspects of the shape of the tube, as we will see in this section. The parallel dynamics are a different matter. The strong compressions of the plasma due to the shortening of the tube generate GDSs along the tubes that present discontinuities in the state variables \citep{Guidoni_2010}. 

To study the full dynamical solution of the TFT equations, we developed a computer program called Dynamical Evolution of Flux Tubes (DEFT) which solves the TFT Equations \eqref{eqn:TFT_mom} and \eqref{eqn:TFT_Pc_time} in dimensionless form for the two reconnected tubes (DOWNWARD and UPWARD moving). Mass conservation is automatically satisfied since the DEFT program implements a Lagrangian approach where each mass element of the tube is followed. The program uses a one-dimensional staggered mesh where each tube piece is represented by grid points at its ends. A rather detailed description of this computer program was presented in Guidoni10.

In this paper, we will present three different simulations that correspond to the three reconnection positions and angles described in section \ref{sec:JES}. The initial tube shapes correspond to a half reconnection angle $\zeta_{R}=45$\degree. Results depend quantitatively on the reconnection angle, but we chose this intermediate value to show the general aspects of the solutions that are common for a wide range of angles. The half angle that field lines make at the center of the CS is $\zeta_{0} \simeq 57$\degree\ for the BOTTOM and TOP cases, and $\zeta_{0} \simeq 45$\degree\ for the CENTER case. 

The central initial part of the tubes are clearly out of equilibrium due to the sharp angle at the reconnection site. To avoid introducing length scales at the limit of resolution, the central part of the initial tube is smoothed. The end points of the tube are assumed to be fixed, and to have no temperature gradient (this last condition ensures no heat transfer from the end points). 

For concreteness, we will present the results from all simulations in units that assume a magnitude of the background magnetic field at the reconnection site of \realB, an initial uniform electron number density of $n_{e} =$ \realnumdens, and initial uniform temperature $T_{e}$ = \realTemp. The Alfv\'{e}n  speed $v_{aeR}$ at the reconnection site is then approximately $2200$ km/s. As in the above section, the half length of the CS is $L_{e} =$ \realL. We assume $\overline{m}$ equal to half the mass of the proton. Each tube has $1100$ points. 

The unitless numbers for this simulation correspond to plasma-$\beta \simeq $ \realbeta, Prandtl number (ratio between the viscosity and thermal conductivity)  $P_{r} = $ \PrandtlSim, and viscous Reynolds number $R_{\eta} = L_{e} v_{aeR} \rho_{e} / \eta \simeq $ \realReynolds. These values are typical values for the high solar corona. 

The simulated evolution of a DOWNWARD moving tube reconnected at the CENTER site is shown in the right panel of Figure \ref{fig:CENTER_JES_sim}. This simulation has similarities with the corresponding JES (left panel). For each given time, the tube consists of three sections: two unperturbed ones on its sides and a central one that moves downward. Times shown were chosen to coincide with the same bend positions as in the corresponding JES (circles are located in the same positions). The concavity of the central portion has the same sign as the JES, although the JESs are more curved. 

The flattening of the center part of the tube is due to the perpendicular component of the viscous force density in Equation \eqref{eqn:TFT_mom}. In the figure, the small crosses near the center of the tube correspond to the location of sub-shocks, where parallel velocities have large gradients and viscosity becomes relevant. At those locations, the perpendicular viscous force acts in the direction opposite to the local curvature vector; therefore, the total force in the curvature direction is reduced. This increases the curvature of the tube, resulting in a change of curvature sign at the center of the tube. A thorough analysis of the shocks is presented in the next section.

In the simulation, the center of the tube moves with approximately constant speed equal to the $y$-projection of the reconnection Alfv\'{e}n speed, which is of the order of $1500$ km/s. This can be understood, at least for earlier times, by comparison with the Guidoni10 model where this is also the case. The initial tube can be approximated as the initial conditions for skewed uniform magnetic fields (same as in Guidoni10). The magnitude of the local Alfv\'{e}n speed decreases toward the bottom edge of the CS, therefore there are sections of the tube, like the center one, whose perpendicular speed is super-Alfv\'{e}nic. The ratio between some tube elements' perpendicular speed and the magnitude of the local Alfv\'{e}n speed can reach values above $1.1$, although this value is relative since the direction of the background magnetic field is different than the one of the tube. 

In over a minute, the central part of the tube arrives at the lower edge of the CS which marks the beginning of the top of the arcade, located at $y=-100$ Mm. The evolution of the tube after its arrival at the top of the arcade is discussed in detail in Section \ref{sec:Top_arcade}.
 
In the Guidoni10 model, the bends are RDs that move at the Alfv\'{e}n speed. We expect similar behavior in our present case. The positions of the circles in right panel of Figure \ref{fig:CENTER_JES_sim} were theoretically calculated as moving at the local Alfv\'{e}n speed along the tube. They coincide with the actual bend positions in the simulation. They continuously slow down in the $y$-direction as they approach the edge of the CS. Their speed in the $x$-direction, determined by the guide field, is constant. Being faster than the bends in the $y$-direction, the tube's center is the first to arrive at the top of the arcade. This difference in speed is what determines the general upward concavity of the moving part the tube. 

At the bends, the initial stationary plasma is deflected along the bisector of the angle between the two adjacent tube portions (average direction of the curvature force). The resulting speed is Alfv\'{e}nic and has a parallel component toward the center of the tube. These bends, as in Guidoni10, are RDs where only the magnetic field direction and the plasma velocity are changed. Notably, the angle at the bends remains approximately constant during the entire evolution of the tube, and equal to $180 - \zeta_{R}$ degrees (same as in the uniform case).  

The BOTTOM reconnection case simulation (shown in left panel of Figure \ref{fig:BOTTOM_JES_sim}) presents the same characteristics as the CENTER one, but the time it takes for the tube to arrive at the top of the arcade is much shorter ($\sim 16$ s), as the reconnection site is closer to the apex of the arcade.

On the other hand, the TOP reconnection case (right panel, Figure \ref{fig:TOP_JES_sim}) presents some differences. The general concavity sign of the tube changes with time, and this change occurs much later in the evolution of the tube compared to the corresponding JES (left panel). Initially, the bends accelerate in the vertical direction and move faster than the center of the tube, leading to a downward concavity (same as in JES). After the bends reach the center of the CS, they start decelerating as they move toward the top of the arcade. The tube's center accelerates continuously, catching up with the bends after approximately $70$ s, where the tube is almost flat. Meanwhile, the perpendicular viscosity force density increases the curvature near the locations of the shocks, curving the center part in the opposite direction with respect to the general curvature of the tube. For this TOP case, the center of the tube is not the first part to arrive at the top of the arcade. The ratio between some tube elements' perpendicular speed and the magnitude of the local Alfv\'{e}n speed reaches values above $1.7$.

%
%
%
\subsection{Parallel dynamics}
  \label{sec:parall_dyn}

The parallel dynamics along the reconnected tubes presents complex features like shocks and coupled non-linear waves. There is no simple equilibrium counter-part to compare it with, as there was for the perpendicular motion. Nevertheless, its complexity is what makes the parallel dynamics very rich in behavior, and this section is devoted to its description. 

The bends accelerate plasma supersonically toward the center of the tube where they collide and are stopped abruptly. As a result, two strong GDSs are launched outward. \citet{Guidoni_2010} describe in detail their inner structure for the uniform and skewed magnetic field configuration. It consists of one thermal front, where most of the heating occurs, and an isothermal sub-shock, where most of the compression occurs. The thickness of the GDSs increases with time and can be of order of tens of megameters, which is comparable to the size of the entire reconnected tube and larger than the particle's mean free path. The large and small plus signs in the right panels of Figures \ref{fig:CENTER_JES_sim}, \ref{fig:BOTTOM_JES_sim}, and \ref{fig:TOP_JES_sim} represent the beginning (heat front) and the end (sub-shock) of the GDSs, respectively, for each simulation time. 

For the three reconnection cases presented in this paper, the GDSs are ahead of the bends at early times after reconnection. Initially, the shocks develop their inner structure quite rapidly until they achieve a near-equilibrium thickness. Meanwhile, the central part of the tube is increasing in length due to the difference in vertical speed between the center and the bends. Therefore, the central portion eventually becomes larger than the distance covered by the shocks, and the bends move ahead. For the TOP reconnection case, this situation is reversed after the bends move to the lower half of the CS, where they slow down and the shocks overrun the bends again. 

The electron density has a sharp increase at the sub-shocks where plasma is extremely compressed. Here, the second term of the right hand side of Equation \eqref{eqn:density change} is the most relevant to determine the electron density. The top panel of Figure \ref{fig:Center_variables_sim} shows density profiles for the CENTER reconnection case where each solid line corresponds to a different time in the simulation. At early times, the density at the center of the tube is over six times the background density. Even though the theoretical density jump limit for steady state GDSs is $4$, we achieve higher values in our simulations because there is an initial transient over-shoot of density at the center of the tube. The plasma requires a finite time to adjust to the correct steady state shock strength. Density decreases as times goes by, although never reaching the equilibrium value. In the figure, we include an inset to the left that reproduces the $t=60$ s electron density profile, where each point of the graph corresponds to a grid point of the tube. This shows that shocks are well-resolved by the DEFT program. 

Density enhancements by shocks are a common feature in any standard reconnection model; this compression occurs at slow-mode shocks or switch-off shocks, which are close relatives of the gas-dynamic shocks of our paper \citep{Longcope_2010_I}. Therefore, dense outflows are usually expected from reconnection. In our patchy reconnection case, central long-lasting, high-density post-shock regions, or plugs, expand and move outwardly from the reconnection site. Their increase in size can be seen in the right panels of Figures \ref{fig:CENTER_JES_sim}, \ref{fig:BOTTOM_JES_sim}, and \ref{fig:TOP_JES_sim}, as tubes descend. The section of the tubes limited by the small crosses corresponds to the plug. This feature should be observable by an X-ray instrument of sensibility suitable to the emission measure. In fact, bright X-ray features apparently shrinking have been observed \citep{McKenzie_2000, McKenzie_2001}. \citet{Sheeley_2002} also describe some coronal inflows as barely visible density enhancements that move sunward, with dark tails forming behind them. 

Between the bends and the sub-shock, the electron density decreases before increasing again at the sub-shocks. This decrease in density is caused by the lateral expansion of the tube as it moves into regions of lower magnetic field (first term in the right-hand side of Equation \eqref{eqn:density change}), as well as by the expansion of the curved tube in its perpendicular direction, although the latter effect is smaller than the former. In the figure, the positive $x$-positions of the bends are shown as circles for each chosen time. After $60$ s, the plasma density in these regions is less than $70$ percent that of the background density. If observed, this tube would present dark and elongated thin regions (low emission measure) that get darker, descend toward the arcade, and grow in length as the bends move along the legs. This resembles the SADs or sinking columns mentioned in the introduction. By the time the tube arrives at the top of the arcade, it would be mostly dark, and approximately half of the dark region would be cold, and the other half would have a range of higher temperatures as the heat fronts move outwardly. 

The GDSs are strong and the temperature jump across them may exceed an order of magnitude. The CENTER reconnection case's temperature profiles are shown in the middle panel of Figure \ref{fig:Center_variables_sim} for the same times as in the top panel. The temperature across the shocks initially increases with time, achieving values higher than $7$ MK, and later decreases. The large and small cross signs are the positions of the beginning and end of the thickness of the shocks, respectively. The length of the thermal fronts increases with time as they move outwardly from the center of the tube. The inset to the left reproduces the $t=60$ s temperature profile, where each point of the graph corresponds to a simulation point in the tube.  

The thermal pressure gradient in Equation \eqref{eqn:TFT_mom} is crucial for the generation of the shocks. If it were erroneously neglected in this equation, the plasma would continuously pile up at the center of the tube without limit. However, this term is not the only factor in the determination of parallel dynamics. The parallel, inviscid part of Equation \eqref{eqn:TFT_mom} can be re-written in the following way 
%
\begin{eqnarray}
   \label{eqn:par_vel} 
      \frac{D v_{\parallel}}{Dt} & = & \mathbf{v}_{\perp} \cdot \frac{D \widehat{\mathbf{b}}}{Dt} - \frac{1}{\rho} \frac{\partial P}{\partial l},
\end{eqnarray}
%
%
showing two possible sources for the change in tube element's parallel velocity (except perhaps inside the sub-shocks where the viscosity contribution may also be large). The first term on the right-hand side is a fictitious force caused by the change in direction of the unit parallel vector as tube elements move. The solid dark squares in Figures \ref{fig:CENTER_JES_sim}, \ref{fig:BOTTOM_JES_sim}, and \ref{fig:TOP_JES_sim} show the position of a generic tube element as it moves along the tube. 

The parallel velocity profile for the CENTER reconnection case is shown in the bottom panel of Figure \ref{fig:Center_variables_sim}, for the same times as in the other two panels. The positive $x$-positions of the bends are also shown as circles for each chosen time. There, the plasma at rest is abruptly accelerated to super-sonic speeds, therefore the slope of the parallel velocity curve is almost vertical at these points. The same occurs at the bends on the other side of the tube where the inflows are directed to the right (positive speed). Since the angle at the bends remains mostly the same as time goes by, the value of the speed achieved at the bends is also approximately constant. The last time shown in this figure is $60$ s, before any part of the tube has arrived to the top of the arcade. 

The parallel velocity after the bends slightly increases due to a small predominance of the first term in Equation \eqref{eqn:par_vel}. This term also non-linearly couples the parallel dynamics (sound waves) to the perpendicular one (Alfv\'{e}n waves), which introduces dispersion. The oscillations in parallel velocity that follow the bends are not an artifact of the simulation; they are real dispersive waves due to this coupling. This effect was not present in Guidoni10 since the tube segments were straight at all times. These waves are smoothed out by diffusive processes, mostly related to thermal conduction. 

The following slow parallel deceleration coincides with the thermal front positions, where the second term of Equation \eqref{eqn:par_vel} becomes relevant. There, the inflows are decelerated when tube elements encounter the heat front, until they are suddenly stopped at the sub-shocks (sharp decrease in parallel velocity near the center of the tube). In the post-shock region, the flows are reversed in direction with respect to the post-bend velocity.  

The BOTTOM reconnection case presents the same general features as the CENTER reconnection case for the three quantities plotted in Figure \ref{fig:Center_variables_sim}. These two cases are similar because their dynamics corresponds purely to the lower part of the CS where the background magnetic field decreases monotonously toward the edge of the CS. 

The TOP reconnection case presents some differences with respect to the other cases. The three panels of Figure \ref{fig:TOP_variables_sim} show the same type of graphs as in Figure \ref{fig:Center_variables_sim}, for the TOP reconnection case. The post-shock electron density (almost one order of magnitude higher than the ambient density) increases while the plug is in the top part of the CS (the tube is being squeezed by the ambient pressure) and decreases after crossing the center of the CS. The squeezing increases the electron density after the bends while those regions are in the top half-plane of the CS, and a pronounced decrease in density occurs after the bends cross the center of the CS. Those regions below the center of the CS expand as the ambient pressure decreases. The decrease in density after the bends can be half that of the background density by the time the tube arrives at the arcade. As the tube descends, a very dense and hot central region develops and moves downward toward the top of the arcade, while very dark and extended regions grow in length to its sides.  

The maximum post-shock temperature is high, over $16$ MK. The thermal fronts are almost as long as the entire tube, as shown in the middle panel of the figure. There, the fictitious force is several times larger than the pressure gradient force, and has opposite sign. Plasma is continuously being accelerated toward the center of the tube after the bends, and stopped at the sub-shock. For a given tube element, it is possible to roughly estimate the value of the fictitious force, for a short period of time, as $\langle \left| \mathbf{v}_{\perp} \right| \rangle \left|\Delta \theta \right|$, where $ \langle \left| \mathbf{v}_{\perp} \right|  \rangle $ represents the average magnitude of the perpendicular velocity in the considered period of time, and $\Delta \theta$ is the corresponding angle change in the parallel unit vector. We have estimated this term for this case, and it is several times larger than the corresponding pressure gradient term in Equation \eqref{eqn:par_vel}. Here, not only the perpendicular velocity is higher, but also the change in angle of the unit vector as tube elements move is larger. 

The speed achieved at the bends is almost unchanged over the entire simulation as the angle at the bends does not change considerably through the simulation. We have run simulations with different reconnection angles and this angle seems to always remain approximately equal to $180 - \zeta_{R}$ degrees. The post-shock parallel velocities are reversed with respect to the direction of the post-bend velocities, and the resulting speeds are high, almost equal in magnitude with respect to the speeds achieved at the bends.

A zoom near the bends in the right panels of Figures \ref{fig:CENTER_JES_sim}, \ref{fig:BOTTOM_JES_sim}, and \ref{fig:TOP_JES_sim} would show small oscillations in the shape of the tubes. There, the fictitious term is much larger than the second term in the right-hand side of Equation \eqref{eqn:par_vel}, as the RDs do not change the pressure of the plasma. 

\subsection{Top of the Arcade}
  \label{sec:Top_arcade}

In the previous sections, the analysis of the evolution of the tubes was done for times when the tubes were above the edge of the CS. The DOWNWARD moving tubes will arrive at the top of the arcade after a finite time. For example, the middle point of the CENTER reconnection tube arrives at the top of the arcade shortly after a minute of travel. This arrival is usually referred in the literature as the place of the fast-mode shock or termination shock \citep{Forbes_1983, Forbes_1986, Forbes_1986_I}. Here, the reconnected flux tubes collide with the underlaying arcade. 

It is likely that the parallel dynamics will continue evolving along the legs of the tube after the tubes had arrived to the top of the arcade. The legs extend downward to the solar surface and do not lay in the same plane of the CS, therefore the simple two-dimensional analysis we have done so far for the evolution of the tube cannot be applied. An exact analysis of the dynamics of the tube after the tube arrives at the edge of the CS escapes the scope of this paper. Nevertheless, we can make some approximations to study it. The decrease in length of the tubes after they arrive at the top of the arcade is probably much smaller than the decrease they undergo since reconnection, as the underlying arcade prevents them from shortening. By the time the tubes have arrived to the top of the arcade, their total lengths decreased by $7$, $15$, and $28$ \%\ for the BOTTOM, CENTER, and TOP reconnection cases, respectively. They may also change their length, only by a small fraction, by decreasing the angle between the legs of the tube and its flat central part (panel (a) of Figure \ref{fig:Top_arcade} shows a cartoon where a reconnected tube (thick line) is laying on top of the underlaying arcade).

To simplify the analysis, we can assume that this length is fixed after the arrival at the arcade apex. The perpendicular motion of the tube is assumed to have stopped, and only the parallel dynamics along the legs would continue, mostly independently of how curved the tube is. To study this parallel dynamics, we will approximate the tube legs as straight line extensions on the sides of the tubes at the edge of the CS and parallel to the CS plane. Therefore, in our simulations an extra straight segment is added to the end of the tubes to simulate the legs; then the tubes are extended in the horizontal axis near the edge of the CS, as shown in the left panel of Figure \ref{fig:CENTER_JES_sim} (dashed lines). 

Most of the observed dark voids slow to a stop when they arrive at the top of the arcade \citep{Sheeley_2004}. \citet{Linton_2009} show through three-dimensional MHD simulations that reconnected flux tubes decelerate rapidly when they hit the Y-lines of a skewed Sirovatski\v{i} CS and the sheared arcade beyond them. Therefore, we chose in this first analysis to ignore the possibility of full ``bounce'' back of the tube toward the reconnection site. We propose a scenario where tubes are slowed down and stopped there by a damping force. We simulate this effect as a restoring force, as shown in panel (c) of Figure \ref{fig:Top_arcade}. The underlying arcade gets compressed and lightly curved due to the arrival of the reconnected tube, and these stronger field lines exert an upward force on the tube. We add a perpendicular damping and spring force densities to Equation \eqref{eqn:TFT_mom}, active only below the edge of the CS
%
\begin{eqnarray}
   \label{eqn:Damp_force} 
      \mathbf{f} & = & \rho \widehat{\mathbf{n}} \left(   \Omega^{2} y - \nu  \widehat{\mathbf{n}} \cdot \mathbf{v} \right) \hbox{   for } y^{\prime}  < -1. 
\end{eqnarray}   
%
Here, $\widehat{\mathbf{n}}$ indicates a unit vector perpendicular to the tube, $\Omega^{2}$ is the spring constant, and $\nu$ is the damping coefficient. This force only acts in the perpendicular direction, slowing down the tube and stopping it. 

The above force density will result in overdamped oscillations when the ratio between the damping coefficient and the spring force, $\xi = \nu / 2  \Omega$, is larger than one. We opted for this in our simulations in Section \ref{sec:simul}. The cases $\xi = 1$ and $\xi < 1$ correspond to critical and under-damped oscillations, respectively. 

Different arcade damping coefficients lead to different damping times, but do not considerably affect the parallel dynamics of the tube. We ran simulations with coefficients differing by orders of magnitude and all results were similar. For the results presented in this paper, the simulations were carried out with coefficients equal to $\xi = 2.0$ and $\Omega \simeq 0.49$ s\textsuperscript{-1}. 

In our simulations, the reconnected tubes retract from the reconnection site to the top of the arcade, and slow down due to the damped spring force from the arcade, until they lay flat on top of it. Then, the final shape of a DOWNWARD moving tube would be a straight line on top of the arcade (see right panels of Figures \ref{fig:CENTER_JES_sim}, \ref{fig:BOTTOM_JES_sim}, and \ref{fig:TOP_JES_sim} for late times in the simulation). 

Since the bends decelerate in the lower part of the CS, they are not the first sections of the tube to arrive at the top of the arcade. The first regions to do so become flat and grow in size as more tube elements descend atop the arcade (compare, for example, $t=65$ and $70$ s of the CENTER simulation in Figure \ref{fig:CENTER_JES_sim}).  We will call the outer edges of these flat regions ``second bend''. 

Some of the parallel velocity change is due to the perpendicular dynamics captured in the first term in the right-hand side of Equation \eqref{eqn:par_vel}. After a section of the tube arrives at the top of the arcade, its perpendicular velocity is suddenly decreased by the damping spring force. Therefore, the first term of Equation \eqref{eqn:par_vel} becomes unimportant compared to the other terms in the equation. The top panel of Figure \ref{fig:Center_rarefaction} shows parallel velocity profiles for late times in the CENTER reconnection simulation (only the positive $x$-axis side). The dashed line corresponds to simulation time equal to $60$ s -the earliest time shown in this graph and the last time shown in Figure \ref{fig:Center_variables_sim}, just before any part of the tube arrives at the arcade. The sharp discontinuity on the left shows the deceleration at the sub-shock. The next times shown in the graph present a new discontinuity at the locations of the second bend, where the parallel velocity is sharply decreased by the sudden reduction in perpendicular speed. The ``SB'' arrows point toward the second bend locations at $t=65$ s and $t=70$ s. 

When the second bend and the original bend become close (see arrow labeled ``SB-B'' in Figure \ref{fig:Center_rarefaction}), the angle at the latter one decreases until it finally becomes $180$\degree\ when the bend arrives at the top of the arcade.  Then, the inflow speed is drastically reduced because it depends strongly on this angle. The gas becomes rarefied and the electron density decreases rapidly behind the bends. Here, previously accelerated plasma continues moving toward the center of the tube leaving a density depletion behind. By the last time of our simulation, the electron density decrease is more than $40$ percent of the background density. The bottom panel of Figure \ref{fig:Center_rarefaction} shows the electron density evolution for the same times shown in the top panel. The almost vertical slopes to the left indicate the sub-shock positions that continue moving outwardly, even after the bends arrive at the top of the arcade. For the TOP reconnection case, the maximum electron density decrease achieved by our simulations is more than $50$ percent of the background.

The sub-shocks will continue moving outwardly and interact with the vacated regions. Although our simulations stop before this occurs, we hypothesize that the electron density will continue decreasing for $|x| < |x_{bottom}|$ and the central hot plug will mix with these rarefaction waves, generating secondary rarefaction waves that would move toward the center of the tube and ultimately will disassemble the hot plug, as described by \citet{Longcope_2010}.  

\subsection{Emission Measure and Mean Temperature}
  \label{sec:rarefaction}

The differences in dynamics between each reconnection case suggest an observational signature of where in the CS a flux tube had been reconnected. The tube's average temperature, weighted by the differential emission measure per flux, DEM(T), can be calculated as
%
\begin{eqnarray}
   \label{eqn:mean_T} 
      \langle T \rangle & = & \frac{ \sum\limits_{T>T_{e}} DEM(T) \hbox{ } T \hbox{ } \Delta T}{\sum\limits_{T>T_{e}} DEM(T) \hbox{ }  \Delta T}, 
\end{eqnarray}   
%
where the DEM(T) is computed for small temperature bins of size $\Delta T$ along the tube, as
%
\begin{eqnarray}
   \label{eqn:DEM} 
      DEM(T) & = & \frac{\sum\limits_{\left |T(l) - T\right| <\frac{\Delta T}{2}} EM(l)}{\Delta T}. 
\end{eqnarray}   
%
Only tube elements whose temperature is inside the range of the bin are included in the sum. $\langle T \rangle$ refers to the heated material of the tube; it excludes segments at or below ambient temperature. Figure \ref{fig:Center_EM_T_evol} shows the time evolution of the tube's average temperature (dotted line) for the CENTER reconnection case.  

The emission measure per flux of a given tube element, EM(l), is computed in the following way
%
\begin{eqnarray}
   \label{eqn:EM_l} 
      EM(l) & = & \frac{n^{2}(l) \hbox{ }  \delta l} {B_{e}(l)}. 
\end{eqnarray}   
%
With the above definition, the tube's total emission measure per flux becomes
%
\begin{eqnarray}
   \label{eqn:EM_flux} 
      \frac{EM}{\phi} & = & \sum\limits_{T(l)>T_{e}} EM(l). 
\end{eqnarray}   
%
The above quantity considers only the emission measure of the heated parts of the tube ($T > T_{e}$), excluding cool sections like the ones following the bends and preceding the heat fronts, shown in the middle panel of Figure \ref{fig:Center_variables_sim}. Ambient plasma emission measure is not included either.  Figure \ref{fig:Center_EM_T_evol} shows the time evolution of the tube's total emission measure per flux (dashed line) for the CENTER reconnection case.  

In the figure, the arrow labeled ``C'' indicates the time at which the center of the tube arrives at the top of the arcade, and the one labeled ``B'' represents the arrival of the bends at the top of the arcade. Initially, the mean temperature of the tube increases at a faster rate than the emission measure. They both achieve a local maximum before time ``C''. The following negative slope for both curves indicates that these quantities would rapidly decrease if the tube were left to retract without the arcade interrupting its motion. Remarkably, they increase after this point. The length of the tube does not change considerably while laying on top of the arcade, but the heat fronts continue moving along the tube, increasing its average temperature. Our simulations end before the heat fronts arrive at the footpoints, but we hypothesize that the average temperature of the tube will continue increasing until they do so.

The central hot plug lasts longer than a conductive cooling time of a region with similar temperature and size \citep{Longcope_2010} as it is maintained by the inflows. These quantities are therefore long-lasting. The TOP reconnection case presents some differences. The mean temperature and emission temperature grow together, as shown in Figure \ref{fig:TOP_EM_T_evol}. Here, the arcade also prevents a decrease in both quantities.

In order to provide a general method to determine where in the CS reconnection happened for a given flux tube, we normalized the mean temperature and emission measure by their maximum achieved value. Figure \ref{fig:T_EM_relation} shows the dependence of normalized emission measure with normalized mean temperature for each of the reconnection cases. The BOTTOM and CENTER case (dotted and dashed lines, respectively) are very similar to each other. For these cases, the mean temperature achieves its maximum before the emission measure does. On the other hand, the TOP case presents a different dependence. Both quantities initially increase simultaneously and curl at the top end of the curve.

\section{Discussion}
  \label{sec:conclusions}

We have presented a model of tube dynamics, accompanied by simulations, following transient and localized magnetic reconnection in a realistic coronal background configuration (Sirovatski{\v i} CS). We have shown, that retracting reconnected flux tubes may present elongated regions devoid of plasma, as well as long lasting, dense central hot regions. The latter are created by GDSs at the center of the tube, consisting in long thermal fronts followed by an isothermal sub-shock. In general, the jump in density across the shock exceeds the maximum value predicted by Rankine-Hugoniot \citep{Rankine_1870,Hugoniot_1887} conditions. These jump conditions are calculated assuming the shock is in steady state, which is not the case here. For the TOP reconnection case, the jump in density can be almost an order of magnitude. This descending plasma plug, although very thin, would be extremely bright. It is long lived compared to a free expansion of similar temperature and size region since it is maintained by the inflows generated at the bends.     

GDSs are also present in reconnected tubes sliding through CS with uniform skewed fields \citep{Guidoni_2010}. However, in our present model, these hot plugs respond to the change in background magnetic pressure that compresses or expands them laterally, depending on which region of the CS the tube is sliding through. For Sirovatski{\v i} CSs, the magnetic field decreases toward the edges of the CS (Y-points). There, its magnitude is minimum (it is zero if there is no guide field). Therefore, it is possible for the tubes to move through regions of decreasing magnetic field as they descend toward the Sun. This seems counter-intuitive because usually magnetic fields in the corona are assumed to increase at lower heights. 

We have presented only three reconnection locations with a reconnection half angle of 45 \degree, for illustrative effects. We used generic dimensional values, but results scale with $B_{e}$, $\rho_{e}$, and $L_{e}$. For all the cases presented, the maximum plasma-$\beta$ achieved never exceeded unity.

RDs (bends) move at the local Alfve\'{e}n speed along the legs of the tubes. Elongated low density regions are generated by lateral expansions behind the bends, when tubes move through the lower half plane of the CS. There, background magnetic pressure decreases toward the edge of the CS. The amount of density depletion depends on how curved the tube is, which is directly related to the perpendicular gradient in the background magnetic field (reconnection angle). On the contrary, in uniform background fields, as shown in Guidoni10, tubes present only straight sections, regardless of the reconnection angle. No density depletion is expected in this case. When there is no significant density depletion, and the GDSs are allowed to extend for a detectable length, the only observed signature of these tubes would be a descending hot plug with high density. 

The achieved decrease in density could be as much as $30$ \%\ to $50$ \%\ of background (pre-flare) values. This level of depletion agrees with observations \citep{McKenzie_1999, Sheeley_2002}. These percentages are lower bounds since we have not considered any increase in the surrounding plasma density by, for example, chromospheric evaporation from previous reconnection episodes. 

\begin{figure}
   \centering
   \begin{center}$
      \begin{array}{cc}
         \includegraphics{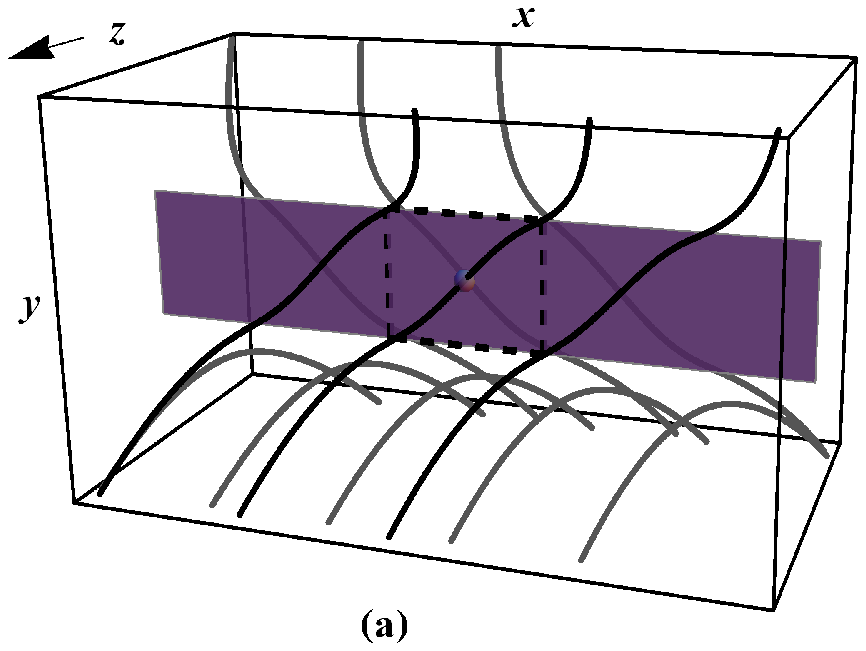}
         \includegraphics{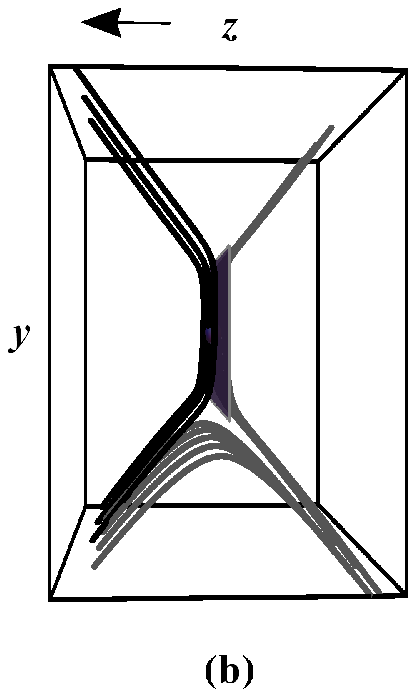}
      \end{array}$
   \end{center}
   \caption
   {
      Flare current sheet geometry. Panel (a): field line configuration for a Green-Syrovatski{\v i} current sheet with a guide field in the positive $x$-direction. The transparent gray rectangle represents the current sheet in the $x-y$ plane, located above the post-flare arcade (gray parabolic loops at the bottom). The current sheet is finite in the $y$-direction, but extends infinitely in the $x$-direction (2 1/2 dimensional symmetry). The bottom plane of the box represents the solar surface. Black lines depict some magnetic field lines on the side of the current sheet that is closer to the viewer ($z > 0$), and the gray ones whose endpoints touch the top side of the box correspond to field lines on the back side of the current sheet ($z < 0$). The small sphere in the current sheet shows a generic patchy reconnection region. Field lines that intersect this region form a small bundle that reconnects. The dashed rectangle in the plane of the current sheet corresponds to the region shown in Figure \ref{fig:CS_lines}. Panel (b): different view of the same field lines in Panel (a), showing the double Y-type configuration. }
   \label{fig:flare_CS} 
\end{figure}

\begin{figure}
  \centering
   \includegraphics{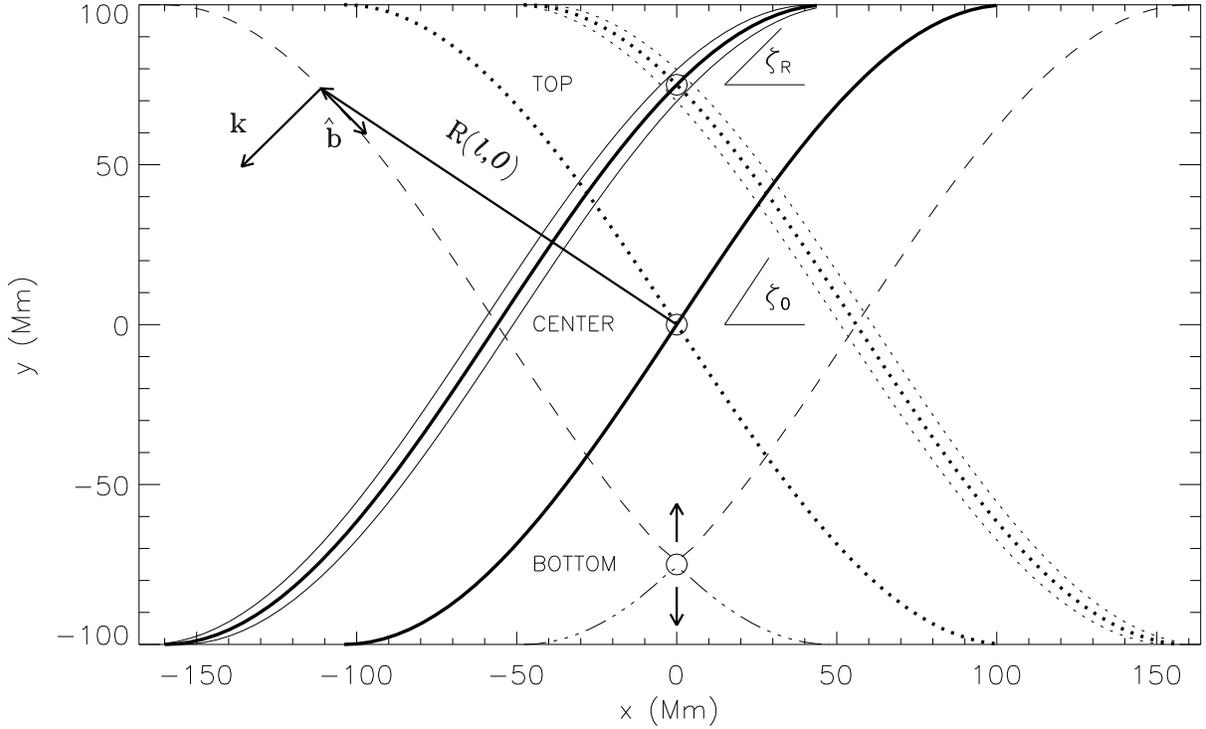}
   \caption{Excerpt of the plane of the current sheet, with three possible reconnection sites. This graph corresponds to the dashed box in Figure \ref{fig:flare_CS}. Pre-reconnection field lines in front of the current sheet ($z > 0$) are represented by solid lines, and pre-reconnection field lines on the back side of the current sheet ($z < 0$) are shown as dotted lines. The edges of the current sheet are located at $y = \pm L_{e} = \pm 100$ Mm. The top circle shows a TOP reconnection site at $y_{R}=\frac{3}{4} L_{e}$. The set of field lines that intersect this region (pre-reconnection tube) is represented by a central line (thicker). At this reconnection site, representative field lines on each side of the current sheet are skewed by $90$\degree. The half reconnection angle ($\zeta_{R} = 45$\degree) is shown to the side of the TOP reconnection circle. The center circle represents a CENTER reconnection site with $\zeta_{R} \simeq 57$\degree. For this site, only the pre-reconnection representative field lines are drawn. The circle in the bottom half of the CS plane, located at $y_{R}= - \frac{3}{4} L_{e}$, represents a BOTTOM reconnection site with $\zeta_{R} = 45$\degree. Here, the initial shape of the already reconnected tubes are drawn, and arrows indicate their direction of motion. The dotted and dashed line shows a reconnected tube that will move downward (DOWNWARD moving tube) due to magnetic tension at its center, and the dashed reconnected tube will move upward (UPWARD moving tube). For this last tube, a parametrization vector $\mathbf{R}(l,0)$ is shown for a given arc-length $l$. Also at this arc-length, the unit vector $\widehat{\mathbf{b}}$ is shown, as well as the curvature vector $\mathbf{k}$. For all three reconnection cases, $\zeta_{0} \simeq 57$\degree, therefore field lines on each side of the current sheet are parallel to each other for all reconnection positions.}
   \label{fig:CS_lines} 
\end{figure}

\begin{figure}
  \centering
   \includegraphics{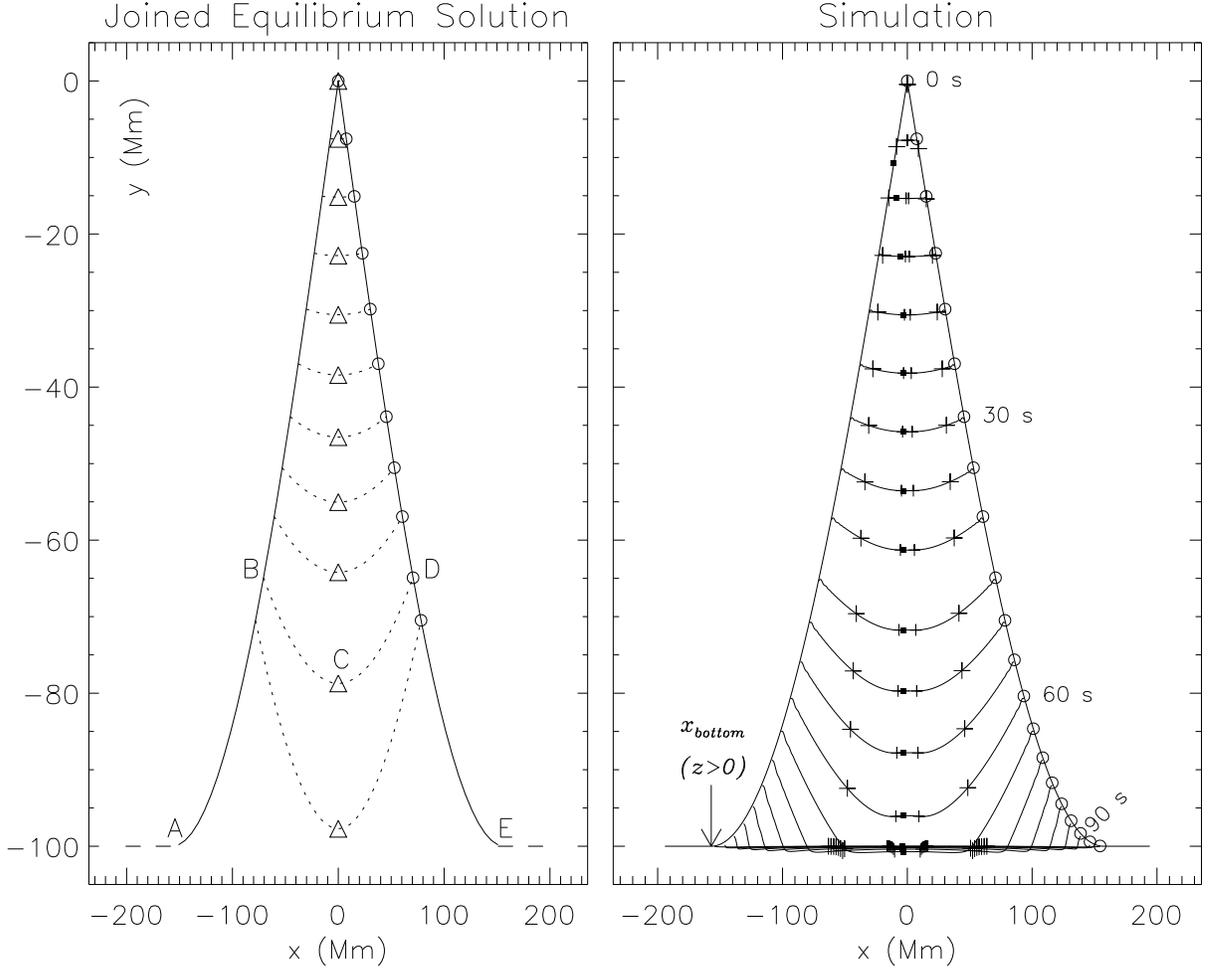}
   \caption{DOWNWARD moving tube for the CENTER reconnection case and reconnection angle $\zeta_{R}=\zeta_{0}=45$\degree\ (note that $y$-axis and $x$-axis are not drawn to scale). Both panels share the same $y$-axis. Left panel: solid line represents the shape of the reconnected tube at $t = 0$. Each dotted line corresponds to a different JES with center position parameter $y_{c}^{\prime} = 0.0,-0.08,-0.15,-0.23,-0.30,-0.38,-0.47,-0.55,-0.64,-0.79,$ and $-0.98$ (shown as triangles), respectively. Circles indicate the bends for the right side ($z < 0$) of the tube. Letters ``ABCDE'' join a complete JES curve for $y_{c}^{\prime} = -0.79$. The horizontal dashed lines show extensions of the tube that simulate the legs that connect it to its footpoints at the photosphere. Right panel: DEFT simulation of the same initial reconnected tube as in left panel. Each central curved line represents the shape of the tube at a different time. The time interval between each curve is approximately five seconds. Circles represent the theoretical positions of the bends (assumed to move at the local Alfv\'{e}n speed). Time in seconds is shown for some selected tube configurations to guide the eye. Large plus signs at the center of the tube indicate the beginning of the gas-dynamics shocks (heat front) and the smaller plus signs indicate the end of the shock (sub-shock). The solid black squares indicate the positions of a tube element as it moves along the tube. The arrow points to the location where the tube becomes tangent to the edge of the current sheet ($x_{bottom}$) for the front side of the tube ($z > 0$). To the left of this point, the tube is extended as a straight line to simulate the leg that connects it to the photosphere (it corresponds to the horizontal dashed line to the left in the left panel). A similar straight line is added to the other end of the tube.}  
   \label{fig:CENTER_JES_sim} 
\end{figure}

\begin{figure}
  \centering
   \includegraphics{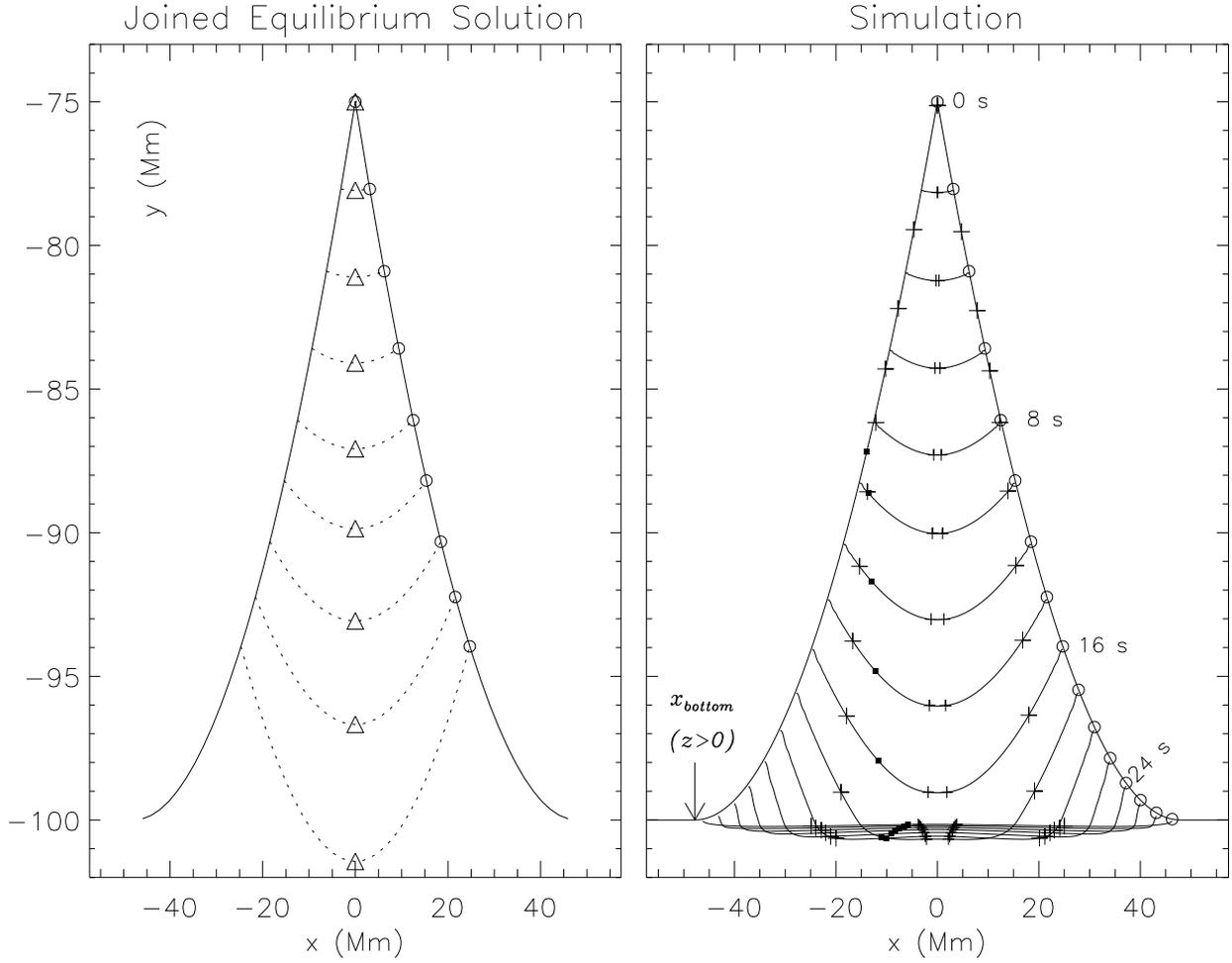}
   \caption{DOWNWARD moving tube for the BOTTOM reconnection case and reconnection angle $\zeta_{R}=45$\degree\ and $\zeta_{0} \simeq 57$\degree, with the same format as Figure \ref{fig:CENTER_JES_sim}. Left panel: the center position parameters are $y_{c}\prime = -0.75,-0.78,-0.81,-0.84,-0.87,-0.90,-0.93,-0.97,$ and $-1.01$, respectively. Right panel: the time interval between each curve is approximately two seconds.}
   \label{fig:BOTTOM_JES_sim} 
\end{figure}

\begin{figure}
  \centering
   \includegraphics{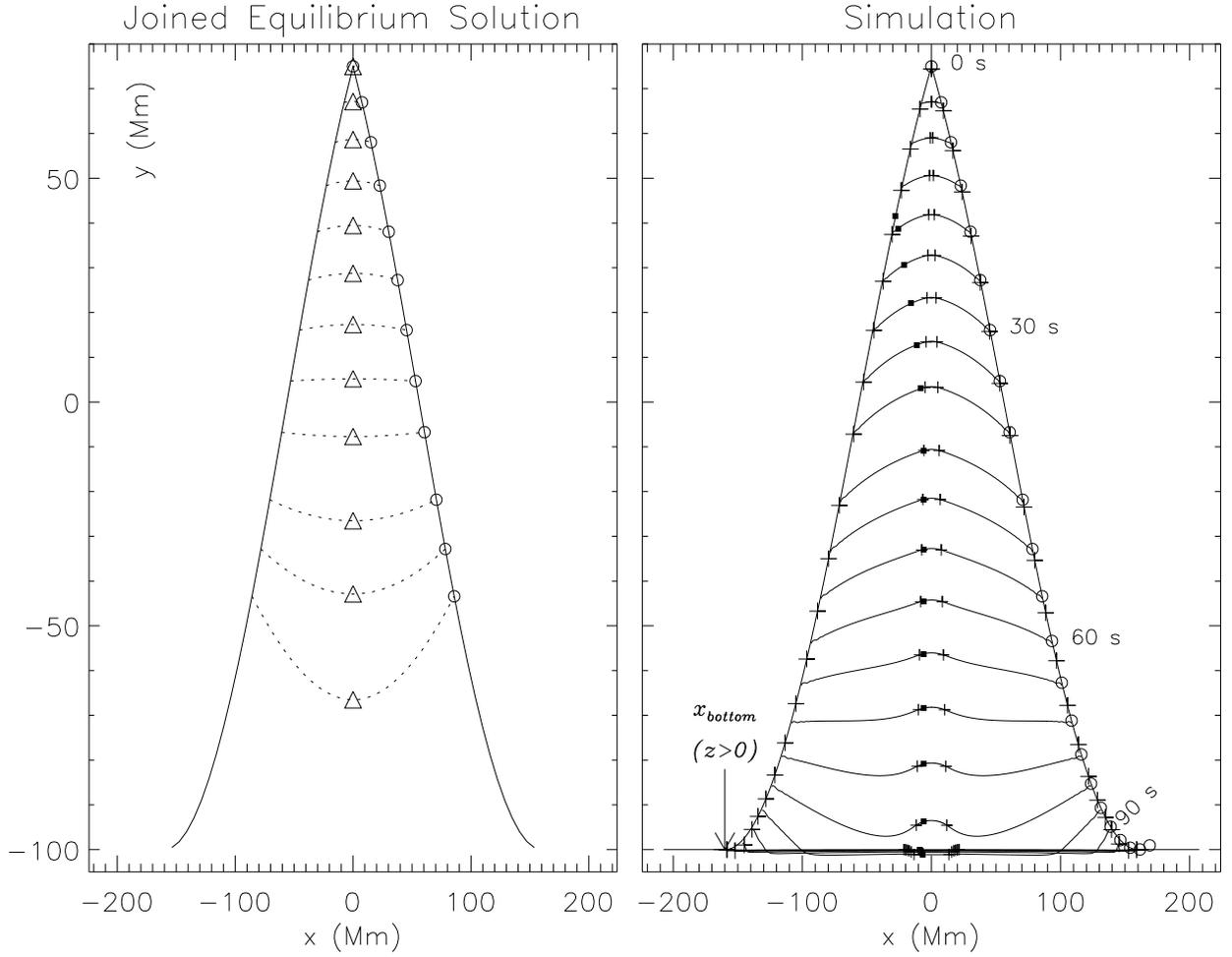}
   \caption{DOWNWARD moving tube for the TOP reconnection case and reconnection angle $\zeta_{R}=45$\degree\ and $\zeta_{0} \simeq 57$\degree, with the same format as Figure \ref{fig:CENTER_JES_sim}. Left panel: the center position parameters are $y_{c}^{\prime} = 0.75,0.67,0.59,0.50,0.40,0.29,0.17,0.05,-0.08,-0.27,-0.43,$, and $-0.67$, respectively. Right panel: the time interval between each curve is approximately five seconds.}
   \label{fig:TOP_JES_sim} 
\end{figure}

\begin{figure}
  \centering
   \includegraphics{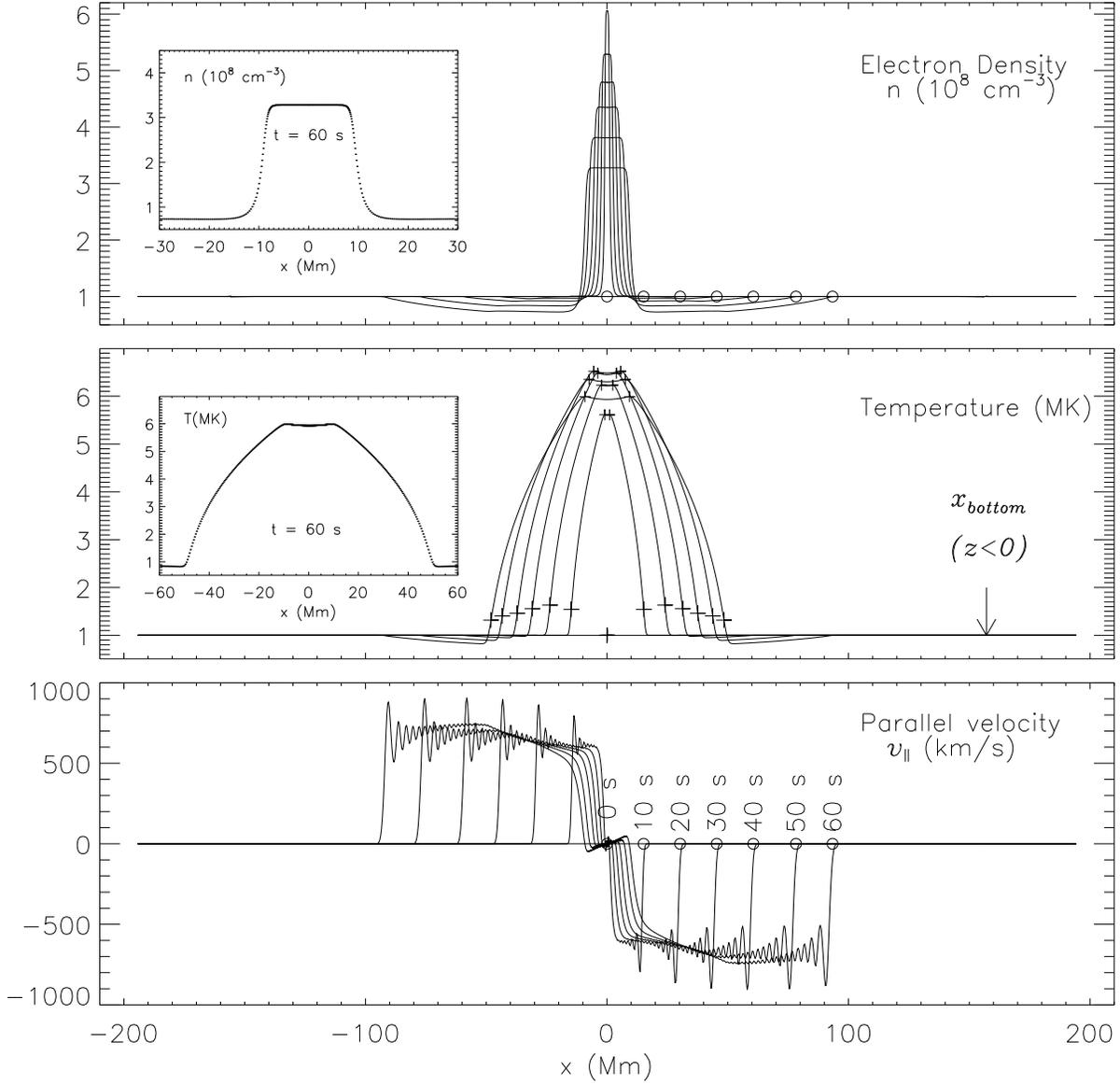}
   \caption{DOWNWARD moving tube for the CENTER reconnection case. All panels share the same horizontal axis. Top panel: electron density along the tube for different times in the simulation. The $x$-positions of the bends (only for the right part of the tube) are shown as circles for each chosen time. Lower densities in the central part of the tube correspond to later times in the simulation. The inset figure repeats the electron density profile for the time $t=60$ s, where each point of the curve corresponds to a grid point of the tube, to show that the sub-shocks are well-resolved by the DEFT computer program. Middle panel: temperature along the tube for the same times as in the top panel. Large and small crosses indicate the beginning (heat front) and end (sub-shock) of the gas-dynamics shocks, respectively. The inset figure repeats the temperature profile for the time $t=60$ s, where each point of the curve corresponds to a grid point of the tube, to show that thermal fronts are well-resolved by the DEFT computer program. The arrow points to the location where the back side of the tube ($z < 0$) becomes tangent to the edge of the current sheet ($x_{bottom}$). Bottom panel: parallel (along the tube) speed profile for the same times as in the top panel. Circles are also the same as in the top panel. Times in seconds are shown for each bend position.}
   \label{fig:Center_variables_sim} 
\end{figure}

\begin{figure}
  \centering
   \includegraphics{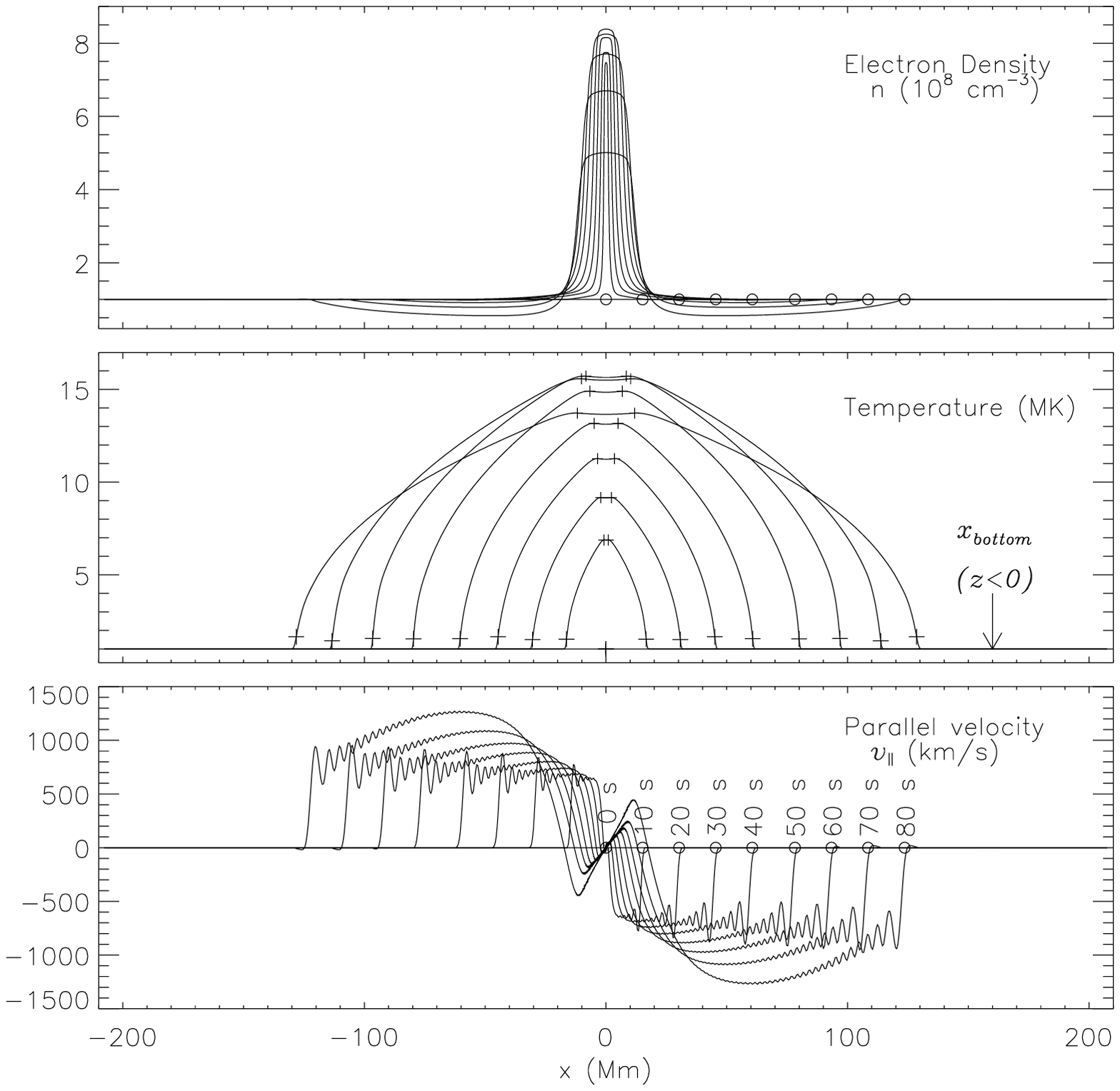}
   \caption{DOWNWARD moving tube for the TOP reconnection case, with the same format as Figure \ref{fig:Center_variables_sim}.}
   \label{fig:TOP_variables_sim} 
\end{figure}

The appearance of the retracting tubes changes considerably with the direction of the line of sight and with the location of the reconnection episode. If reconnected tubes (their observable parts) are seen from the same view as in panel (a) of Figure \ref{fig:flare_CS}, they would have loop-shapes, although their orientation would be perpendicular to the arcade since they move in a plane that is parallel to the arcade axis. On the other hand, if the reconnected tubes are seen from the view shown in panel (b) of Figure \ref{fig:flare_CS}, they would be seen as ``hairpins'' because they are tangent to the plane of the CS. 

For the TOP reconnection case, the concavity of descending loops when they move in the top part of the CS plane is the one that would be expected from a cusp-shaped loop. In this region, these loops would be bright at their sides (they are being squeezed by ambient plasma) and even brighter at their center. If the line of sight is parallel to the $x$-direction, this loop would appear like a bright small region, descending at Alfv\'{e}nic speeds. After the bends cross the center of the CS, dark regions will develop behind them, and this view will change to vertical dark regions preceded by a bright region, until the center catches up with the bends and the situation is reversed. The dark and bright regions are hot since the heat fronts move at speeds similar to the bend speeds. Their sizes are comparable to the entire tube. 

For BOTTOM and CENTER reconnection cases, their concavity is always U-shaped. If seen from the $x$ direction, a bright region is followed by an elongated vertical dark one that might be hot or not, depending on the location of the heat front.  

Observed coronal inflows can be bright or dark, or both; they may also have loop- or tadpole-shapes \citep{McKenzie_2000, McKenzie_2001,Sheeley_2002}. It is possible that many of them are the manifestation of the same three-dimensional phenomena observed from different lines of sight. The location of the tube with respect to the CS also matters. If an observation occurs, for example, while a tube is in the top half-plane of the CS, only bright regions would be observed. On the other hand, if a tube is moving through the lower portion of the CS, it would present dark regions, sometimes preceded by bright regions, or sometimes followed by bright regions, as described above.

A dense, hot region may not always been observed due to its small size, or its high temperature. For reconnection angles larger than $45$\degree, it is possible to achieve post-shock temperatures that are much higher than the usual instrument passbands. On the other hand, after the tube's arrival at the top of the arcade, this region continues growing in time, and may become visible. The pile up of several newly reconnected tubes may also increase the size of this region to observable size, and would manifest as a stationary bright region on top of the arcade.

Remarkably, signatures of reconnection persist in a single tube, longer than the Alfv\'{e}nic transit-time required for the tube to relax. As the tubes move downward, they encounter the top of the arcade that lays beneath the bottom edge of the CS. Here, the downward motion is halted by the arcade. We simulated this effect with a perpendicular damped spring force exerted by the compressed arcade. With this force, the tube comes to rest on top of the arcade as a straight line. The parallel dynamics continues along the legs. After they stop, the tubes cannot decrease their length any longer, and the RDs are shut down. Then, the gas gets rarified even more because the already accelerated plasma continues moving toward the center of the tube, increasing the density depletion behind it. 

Although our simulations stop before this, we hypothesize that the hot post-shock regions will continue moving along the tube and interact with the rarefaction waves, subsequently disassembling the hot plug. The legs of the tube would also brighten up as the thermal fronts descend toward the footpoints.  

The tube's parallel and perpendicular dynamics are non-linearly coupled to each other, which can be seen, for example, in the oscillations of the parallel velocity profiles. In addition, when the tube arrives at the arcade, the perpendicular dynamics is halted, and changes in the parallel speed are evident at the ``second bend''. The parallel velocity profiles are determined by pressure and fictitious forces due to the motion of tube elements along curved paths. This can have important consequences for Doppler-shifts observations. For example, the post-shock region in the TOP reconnection case has very strong flows in a direction almost parallel to the surface of the Sun. This would manifest as coincident blue- and red-shifts along the line of sight, if seen from the $x$-direction, and perpendicular to the line of sight if seen from the $z$-direction. Perpendicular velocities can be higher or lower than the local background Alfv\'{e}n speed, which can be related to the generation of secondary shocks as the tubes move (the surrounding plasma may be shocked by the passing tubes).

 We described the temporal behavior of the total emission measure and mean temperature of the heated parts of the tube and provide an observational method that may indicate where in the CS the tube has been reconnected. Tubes that were reconnected in the TOP half of the CS present total emission measure that grows simultaneously with the mean temperature of the tube. On the other hand, tubes that have been reconnected in the BOTTOM part of the CS achieve a maximum in their mean temperature much earlier than the emission measure does. For all the reconnection cases, the emission measure and mean temperature duration is extended by the arrival of the tubes at the top of the arcade. After the arrival, the length of the tube remains the same but the heat fronts continue heating the plasma as they move along the legs. 

If observed downflows (SADs and SADLs) are reconnection outflows, the question of why their speeds are lower than the assumed Alfv\'{e}n speed still remains. We restricted ourselves to standard reconnection scenarios where the outflows are Alfv\'{e}nic. Nevertheless, if dark voids are related to the dark regions generated behind the bends, their speeds near the edge of the current sheet are decreased considerably from the one they had at the center of the CS. For instance, in the CENTER reconnection case, the bend speeds are reduced by a fourth of what it was at the reconnection site. This decrease in speed of the bends is not enough to explain the observed descending voids that move at half of the presumed reconnection Alfv\'{e}n speed. However, if the half reconnection angle is greater than $60$\degree, the ratio of the Alfv\'{e}n speed at the center of the CS to the edge of the CS is larger than $1/2$. 

\begin{figure}
  \centering
   \includegraphics{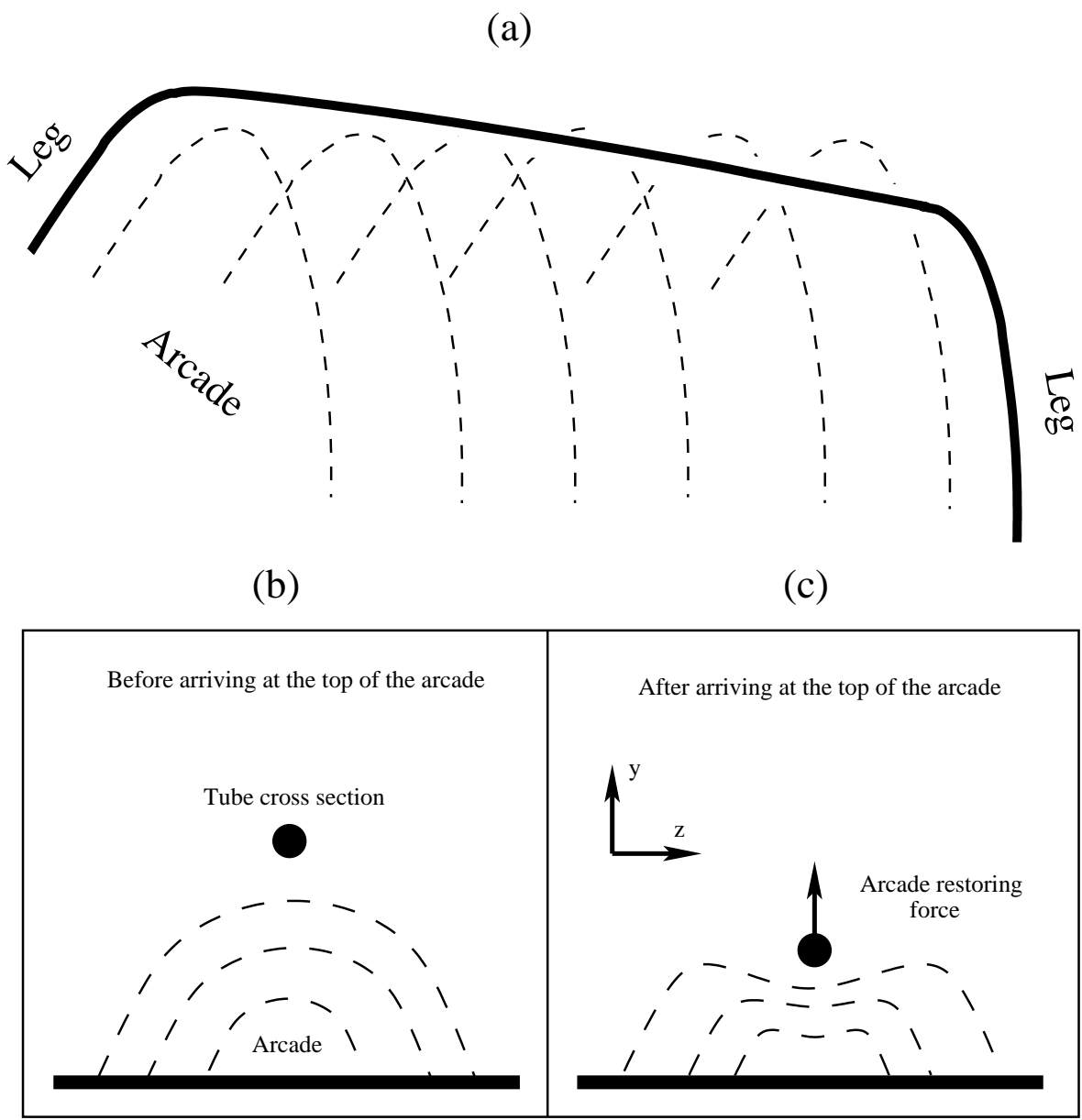}
   \caption{Tube arrival at the top of the arcade cartoon. Panel (a): reconnected tube (dark solid line) lays on top of the arcade after having descended from the reconnection site. Dashed lines represent the underlaying arcade. The legs of the tube connect the parts of the tube that were close to the CS with the tube's footpoints at the solar surface. Panel (b): cross section of the tube (solid black circle) before its arrival at the top of the arcade (dashed parabolic lines). Panel (c): arcade restoring force. The underlying arcade gets compressed and lightly curved due to the arrival of the reconnected tube, and these stronger field lines exert an upward force on the tube. }
   \label{fig:Top_arcade} 
\end{figure}

\begin{figure}
  \centering
   \includegraphics{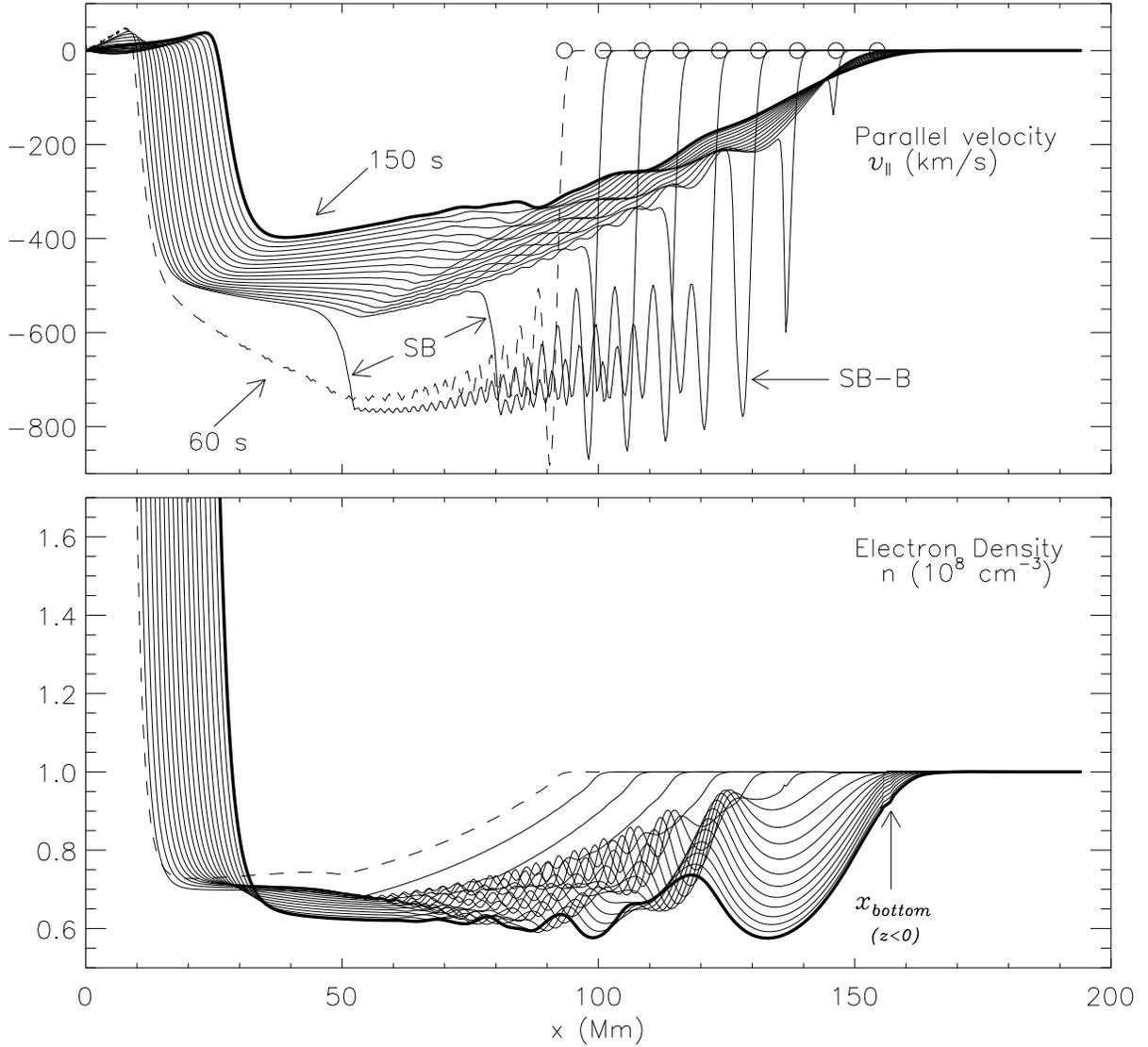}
   \caption{DOWNWARD moving tube for the CENTER reconnection case. The two panels share the same horizontal axis that corresponds to the right side of the reconnected tube. Top panel: the parallel component of the speed for times between $60$ and $150$ s, at five second intervals. The first time (dashed line) corresponds to $t = 60$ s, which is the last time shown in Figure \ref{fig:Center_variables_sim}. Time $t = 150$ s is shown with a thicker line. Circles show the $x$-position of the bends. The ``SB'' arrows point toward the second bend locations at $t=65$ s and $t=70$ s. The ``SB-B'' arrow indicates a time when the second bend and the original bend are close. Bottom panel: electron density profile for the same times as in the top panel. The arrow points to the location where the tube becomes tangent to the edge of the current sheet.}
   \label{fig:Center_rarefaction} 
\end{figure}

\begin{figure}
  \centering
   \includegraphics{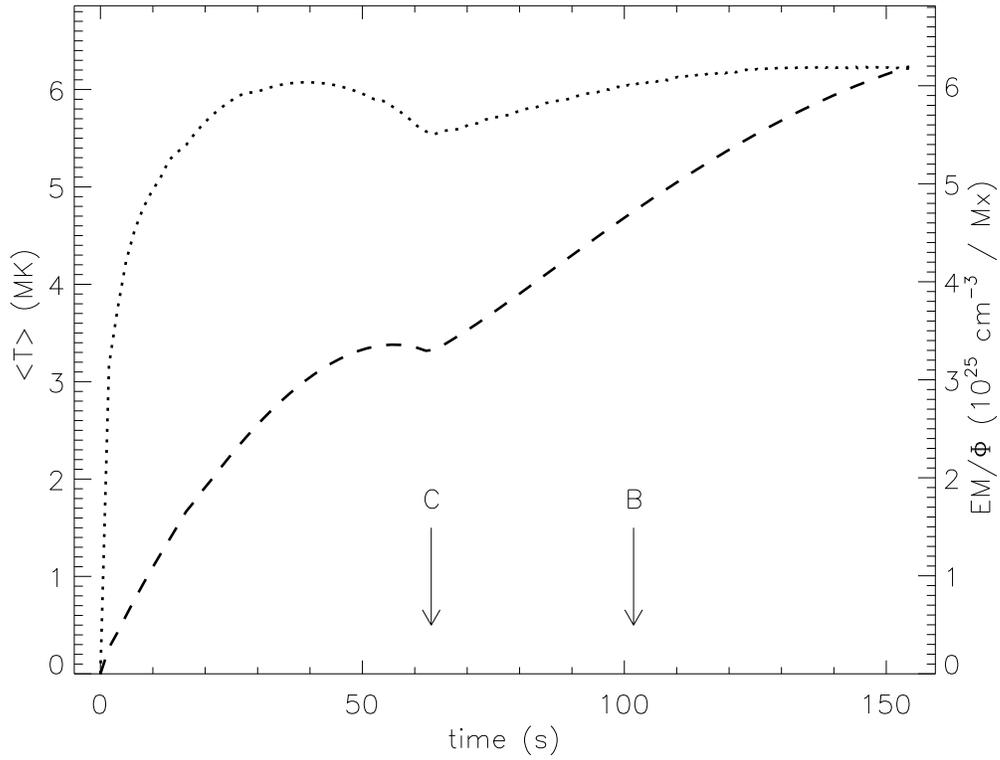}
   \caption{DOWNWARD moving tube for the CENTER reconnection case, with the same format as Figure \ref{fig:CENTER_JES_sim}. The time evolution of the mean temperature of the tube (left vertical axis) is shown with a dotted line. The total emission measure per flux of the tube (right vertical axis) is also shown for the same times with a dashed line. Arrow ``C'' indicates the time when the center of the tube arrives at the top of the arcade. Arrow ``B'' indicates the time when the bends arrive at the top of the arcade.}
   \label{fig:Center_EM_T_evol} 
\end{figure}

\begin{figure}
  \centering
   \includegraphics{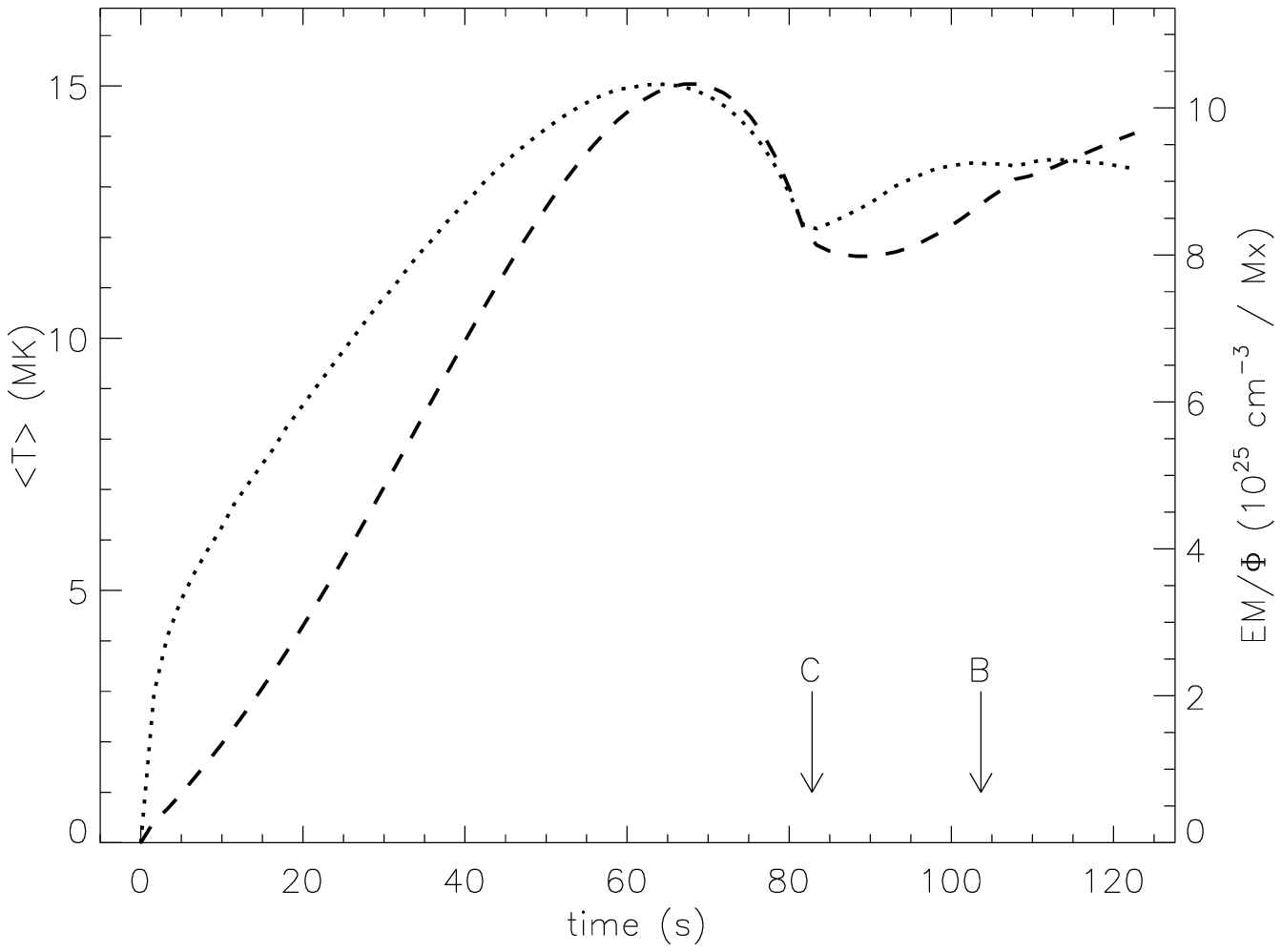}
   \caption{DOWNWARD moving tube for the TOP reconnection case, , with the same format as Figure \ref{fig:Center_EM_T_evol}.}
   \label{fig:TOP_EM_T_evol} 
\end{figure}

\begin{figure}
  \centering
   \includegraphics{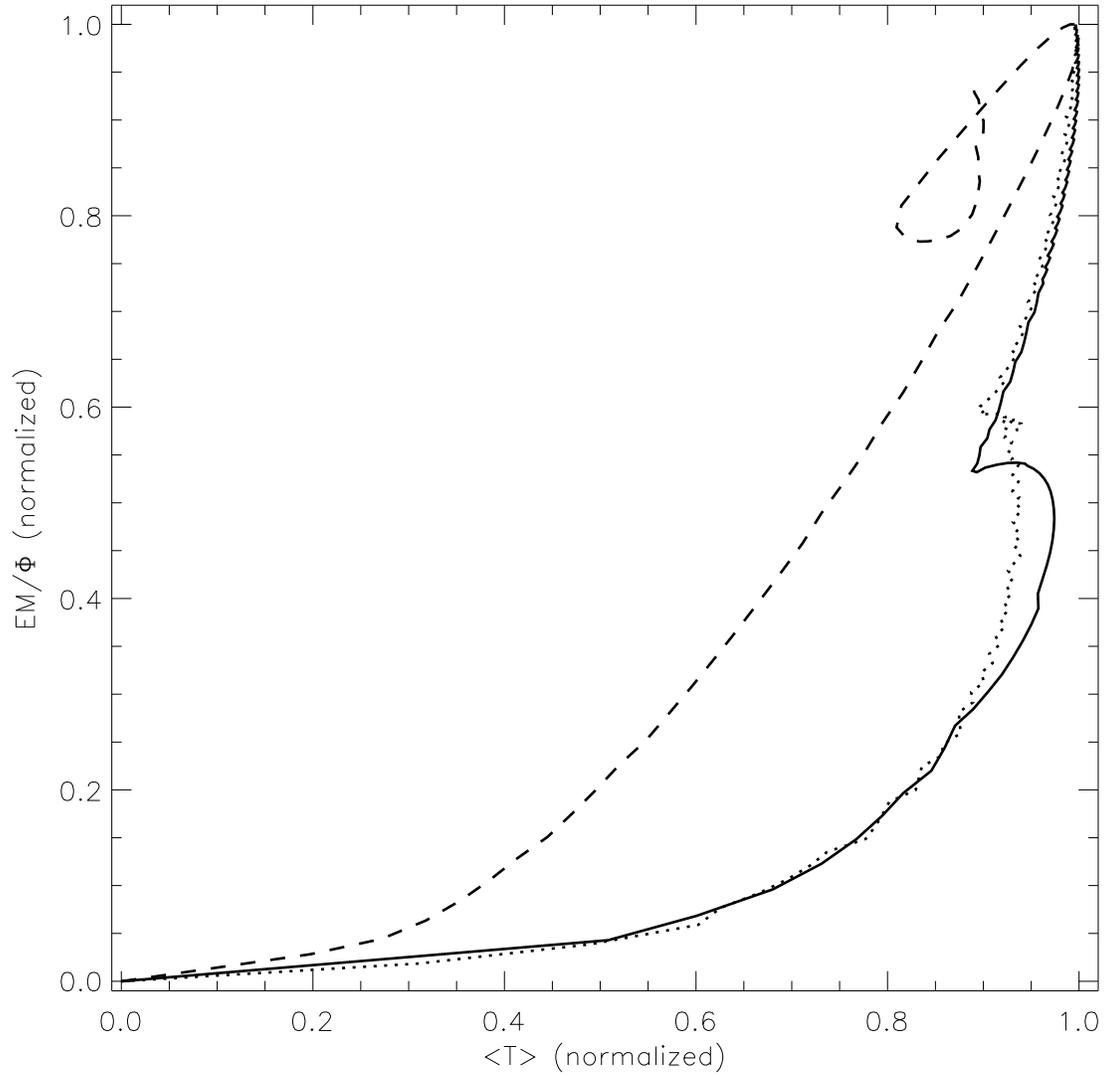}
   \caption{Normalized emission measure as function of the normalized mean temperature of the tube for DOWNWARD moving tubes. The solid line corresponds to the CENTER reconnection case, the dotted line to the BOTTOM reconnection case, and the dashed line to the TOP reconnection case.}
   \label{fig:T_EM_relation} 
\end{figure}

Hard X-ray sources have been observed near top of SXT arcades, at the same time as the SADs descend to the top of the arcade \citep{Asai_2004}. This may be the result of the interaction between downflows and the fast shock, or could be related to reconnection at the arcade apex between the retracting tubes and the arcade. Apex reconnection is a possibility in our three-dimensional model, but one we have not yet explored. Field lines that arrive at the top of the arcade have a different angle than the underlying arcade (panel (a) of Figure \ref{fig:Top_arcade}), and arrive there at Alfv\'{e}nic or super-Alfv\'{e}nic speeds. This is not an option in purely two-dimensional models where the new reconnected field lines arrive at the arcade with magnetic field direction parallel to the arcade.

Maintaining a fast-shock on top of the arcade may not be feasible at sub-Alfv\'{e}nic speeds. The standing fast shocks exist as long as the reconnection jets exists, therefore it is a feature of a flare gradual phase \citep{Forbes_1986_I}, unless reconnection continues during the decay phase. Although some SADs have been reported during this phase \citep{Asai_2004}, most of them were seen during the decay phase. It is not clear if the termination shocks would still be located at the top of the SXT loops if only SADs are the only reconnection outflows. 

Another possibility could be that the termination shock is located further up from the top of the SXT loops. If there are much hotter loops above the SXT loops, that cannot be seen in pass bands at lower temperatures, the termination shock could be located higher in the corona. In this case, the downflows could be localized downstream flows from the fast-shock, which would explain their low speeds. The models for fast shocks are two-dimensional and it is not clear how this would apply to localized and time-dependent flux tubes like the ones described here. The down side to this explanation is that a fast shock certainly would compress the plasma downstream instead of vacating it. 

We have not consider other interactions between the reconnected tubes and their ambient plasma, besides pressure balance between them. The addition of a drag force to our equations could be an important improvement of our model that would contribute to the decrease in speed, as well as to make the model more realistic. We expect to include this effect in future work. The assumption of initial uniform density could also be improved by assuming a stratified atmosphere. Here, tubes that have been reconnected in the corona will descend in regions of higher density, and seen darker than the background, as suggested by \citet{Savage_2010}, although this effect has not been corroborated by theoretical analysis. 

We have neglected any non-fluid effects and therefore neglected non-thermal particles and their acceleration. 

We do not claim that the depleted regions from our model can completely explain the observed dark voids, but it is suggestive that several observed phenomena have some similarities with our model. \citet{Hudson_2001} stated that ``Dark outflows seem inconsistent with the idea of heating by reconnection''. We believe our work demonstrates the contrary, and thus makes the role of reconnection in flares still more plausible.

%
%
%



\clearpage

%
%

\clearpage

%
%

\clearpage

%
%
%

\clearpage

%
%
%

\clearpage

%
%
%

\clearpage

%
%
%

\clearpage

%
%
%

\clearpage

%
%
%

\clearpage

%
%
%

%
%
%
\clearpage

%
%
%

\clearpage

%
%
%

\clearpage

%

\end{document}